# PM2D: A parallel GPU-based code for the kinetic simulation of laser plasma instabilities in large scale plasmas

Hanghang Ma[1,2,3,*], Liwei Tan[4,*], Suming Weng[2,3,*], Wenjun Ying[5], Zhengming Sheng[1,2,3], Jie Zhang[1,2,3]

[1] *Tsung – Dao Lee Institute, Shanghai Jiao Tong University, Shanghai* 200240, *China*

[2] *Key Laboratory for Laser Plasmas* (*MoE*), *School of Physics and Astronomy, Shanghai Jiao Tong University, Shanghai* 200240, *China*

[3] *Collaborative Innovation Center of IFSA, Shanghai Jiao Tong University, Shanghai* 200240, *China*

[4] *School of Mathematical Sciences, Shanghai Jiao Tong University, Shanghai* 200240, *China*

[5] *School of Mathematical Sciences, MOE–LSC and Institute of Natural Sciences, Shanghai Jiao Tong University, Shanghai* 200240, *China*

## Abstract

Laser plasma instabilities (LPIs) have significant influences on the laser energy deposition efficiency, hot electron generation, and uniformity of irradiation in inertial confined fusion (ICF). In contrast to theoretical analysis of linear development of LPIs, numerical simulations play a more and more important role in revealing the complex physics of LPIs. Since LPIs are typically a three-wave coupling process, the precise kinetic simulation of LPIs requires to resolve the laser period (around one femtosecond) and laser wavelength (less than one micron). In the typical ICF experiments, however, LPIs develop in a spatial scale of several millimeters and a temporal scale of several nanoseconds. Therefore, the precise kinetic simulations of LPIs in such scales require huge computational resources and are hard to be carried out by present kinetic codes. In this paper, a full wave fluid model of LPIs is constructed and numerically solved by the particle-mesh method, where the plasma is described by macro particles that can move across the mesh grids freely. Based upon this model, a two-dimensional (2D) GPU code named PM2D is developed. It can simulate the kinetic effects of LPIs self-consistently as normal particle-in-cell (PIC) codes. Moreover, as the physical model adopted in the PM2D code is specifically constructed for LPIs, the required macro particles per grid in the simulations can be largely reduced and thus overall simulation cost is considerably reduced comparing with typical PIC codes. Moreover, the numerical noise in our PM2D code is much lower, which makes it more robust than PIC codes in the simulation of LPIs for the long-time scale above 10 picoseconds. After the distributed computing is realized, our PM2D code is able to run on GPU clusters with a total mesh grids up to several billions, which meets the typical requirements for the simulations of LPIs at ICF experimental scale with reasonable cost.

*These authors contributed to this work.

*Corresponding author. Email: wengsuming@sjtu.edu.cn

# 1、Introduction

Inertial confinement fusion (ICF) is one of the potential approaches to the generation of fusion energy, which is considered as the ideal future energy for the society of human beings due to its characters of being limitless, safe, and clean. The realization of ICF ignition recently on NIF is believed to be a milestone in the development of fusion energy, as well as a huge scientific achievement with important applications in basic sciences [1], [2], [3]. It provides conclusive evidence to the viability of generating fusion energy by laser driven ICF. However, due to the low conversion efficiency from electricity to laser energy (about 1 percent) [4], a much higher energy gain is required in the commercialization of inertial fusion energy. To this end, the efficiency of the laser–plasma coupling as well as the implosion symmetry of the fusion target should be further improved. One of the main reasons that may lead to the implosion asymmetry is the uneven laser energy deposition on the capsule induced by the laser plasma instabilities (LPIs) that are developed during the laser propagation in plasma. Numerical simulations are essential to understand the relevant physics of LPIs, and then to mitigate or even to utilize them. In this paper, we introduce an effective kinetic simulation tool to investigate LPIs under physical scales comparable with ICF experiments, which may help to control or utilize LPIs.

LPIs are important processes for both indirect and direct drive ICF. Key LPIs include stimulated Raman scattering (SRS), stimulated Brillouin scattering (SBS), cross beam energy process (CBET), and two plasmon decay (TPD) instability. As LPIs are essential resonant processes among the electromagnetic waves (light waves) and plasma waves [7], the precise simulation of LPIs needs to resolve the wavelengths of either light wave or plasma waves, which results in a grid size of around several nanometers in the simulations. Notice that in the practical ICF experiments LPIs may be encountered in a spatial scale of several millimeters. Therefore, the required grid meshes can be $10^{11}$ or larger even in 2D simulations [13], which poses a big challenge for the present simulation schemes. As highly nonlinear processes, LPIs are sensitive to the changing of plasma conditions. In particular, the developments of LPIs in the small-scale plasmas could be distinctly different from those in the large-scale plasmas as encountered in ICF experiments [10], [14]. Therefore, the simulation results at small scales cannot fully reveal the evolution rules of LPIs in the ICF experiments. Moreover, the kinetic effects that are responsible for the damping of plasma waves and the generation of hot electrons should also be included in the large-scale simulations of LPIs to improve the simulation accuracy [15]. Namely, the kinetic simulation of LPIs in the large-scale plasma is necessary for the accurate evaluation of the influences of LPIs in the ICF experiments although it is extremely challenging.

At present, there are mainly two kinds of kinetic codes that can be applied to the simulation of LPIs, namely, the particle-in-cell (PIC) codes and Vlasov codes. In general, Vlasov codes require a huge computational cost and thus are mainly applied to the 1D LPIs simulations [16]. By comparison, PIC codes have a faster simulation speed than Vlasov codes and thus are massively adopted in the simulation of LPIs in the kinetic regime. However, the computational costs of the PIC codes are still too large to be used for the 2D or 3D simulations of LPIs in the ignition-scale plasmas [17]. Moreover, PIC codes usually suffer severe numerical noise in the simulation at long-time scales.

To solve this problem, many codes based on the fluid models of LPIs have been developed. The most frequently used codes are based on the enveloped three-wave or five-wave coupling equations using the paraxial approximation, for example, the codes of PARAX [18], NEWLIP [19], PF3D [20], LAP3D [21], HERA [22], and so on. The advantages of these enveloping codes are that they can be applied to simulate the LPIs in the large-scale plasmas as encountered in ICF experiments [7]. However, the application of the paraxial approximation restricts the propagating directions of incident laser beams to a small angle range. This makes these codes hard to be applied to the simulation of LPIs under some practical laser incident schemes, where several groups of laser beams are injected with relatively large intersection angles. Moreover, the enveloping codes generally omit most of the nonlinear and kinetic effects of LPIs, which may lead to a low confidence of the simulation results. In the SBS process, for example, the harmonic generation of ion acoustic waves are a typical nonlinear phenomenon that play important roles in the saturation of SBS. To include this nonlinear effect in the simulation, Hüller *et al.* developed a simulation code named as HARMONY1D/2D for the SBS instability using the harmonic decomposition method. However, kinetic effects such as the hot electron generation are not included in this code.

Actually, these fluid codes based on either paraxial approximation or harmonic decomposition can be regarded as post-processing codes, which adopt the plasma conditions from the radiation hydrodynamics simulations as the initial conditions in their simulations [24]. However, as mentioned above, LPI has remarkable influences on the evolution of radiation hydrodynamics and thus it should be coupled to the hydrodynamics code to obtain a more accurate simulation results. Because the spatial-temporal scales of the radiation hydrodynamic process are much larger than those of the growth of LPIs, incorporating LPIs effects accurately into the radiation hydrodynamics codes poses a big challenge. This problem was partly addressed by incorporating the effects of LPIs into the ray tracing module of the radiation hydrodynamics codes. For example, researchers have coupled the CBET into the radiation hydrodynamics codes to evaluate the energy transfer effects in both direct-drive and indirect-drive ICF schemes [9] [25] [26] [27] [24][20], and good agreements between the simulation results and experimental results were obtained. Recently, Golaitis *et al.* improved the conventional ray tracing model in the PCGO code by replacing needle like rays with thick rays that have Gaussian intensity distribution in their cross sections [25]. The application of thicker rays enables to reproduce the practical laser field with randomly distributed speckles, which will largely improve the simulation accuracy of the in-line CBET model and result in a high confidence simulation results of the PCGO code.

Another kind of codes for the simulation of LPIs are the so-called full wave fluid codes, in which the envelope approximation is abandoned. In this way, some nonlinear effects of LPIs can be included and the corresponding simulation results are thus more reliable. For example, Myatt and Follett *et al.* developed a code named as laser plasma simulation environment (LPSE) for the simulation of LPIs [28]. The LPSE code consists of many packages that are developed for the simulation of different LPI processes including SRS, SBS, CBET as well as TPD [29], [30], [31]. Especially, the TPD module in the LPSE code can be used to simulate the hot electron generation by adopting the test electrons. The LPSE code is a powerful simulation tool due to its capability of simulating LPIs in ignition relevant spatial-temporal scales using billions of grid cells, while including most of the nonlinear effects of LPIs as well. However, the LPSE code handles different LPI processes individually by different packages, and thus it omits the interplay among different LPIs. Moreover, the LPSE code cannot describe the kinetic effects of LPIs self-consistently. Recently, Hao *et al.* developed a full wave code FLAME 1D/MD for the simulation of LPIs [32]. This code is based on a simplified fluid model without using the envelope approximation in both time and space, which makes it suitable for the simulation of nonlinear evolution processes of LPIs at relatively large spatial-temporal scales. However, this code cannot describe some kinetic effects self-consistently, such as the hot electron generation in either SRS or TPD process which is an important influence of LPIs.

Although different kinds of LPI simulation codes with their own unique advantages have been developed in the past few decades, there is still no simulation tool that can be applied to simulate LPIs in the ignition relevant scales while handling with the kinetic effects of LPIs self-consistently. Vu *et al.* proposed to simulate LPIs by the so-called Reduced PIC (RPIC) code based on the Zaharov equations [33]. In the RPIC, the macro particles are advanced to move by fluid forces from the LPI models. As a result, the spatial and temporal resolutions of the RPIC can be largely decreased, which finally lead to several times increase in the simulation speed comparing with the conventional PIC codes. By using the freely moving macro particles, RPIC can also capture the kinetic effects of LPIs. However, several times increase in the simulation speed is not enough for LPI simulation at the ignition relevant scales. As a kind of PIC code, the numerical noise of RPIC may also restrain its applications in the simulations of LPIs at the long-time scales.

In our present study, a full wave fluid model that describes the evolution of several kinds of LPIs including SRS, SBS, CBET as well as the filamentation of laser beams is constructed. This model is solved numerically by using the particle mesh (PM) method in the 2D geometry, and a PM2D (Particle Mesh code in 2 dimension) code is developed. As the macro particles can move freely across the grids, this PM2D code is able to describe the kinetic effects of LPIs self-consistently. Moreover, since this model is specifically designed for the simulation of LPIs like the RPIC code, the number of macro particles allocated in each grid can be greatly reduced comparing with that in the conventional PIC codes. This can reduce the simulation cost greatly, which makes our PM2D code more suitable for the simulation of LPIs at the experiment scales. The differences of our PM2D code from the conventional PIC codes are that the macro particles are pushed by fluid forces in the PM2D code, while they are driven to move by the Lorentz force in the conventional PIC codes. For typical LPI simulations using CPU processors, our PM2D code is about thirty times faster than the conventional PIC code. To further increase the simulation speed, we implement the parallelization of our PM2D code on GPU clusters using the MPI (Message Passing Interface), which makes it possible to simulate the LPIs at ICF experimental scales with a total grid meshes up to several billions.

In the rest of this paper, we will first introduce our physical model adopted in the PM2D code, and then introduce the numerical schemes and parallel strategies in solving the physical model. After that, the benchmark of the PM2D code is given and the typical LPI processes of SRS and CBET are simulated and compared with the results obtained by the conventional PIC code. At the end of this paper, a large spatial-temporal scale simulation is carried out to demonstrate that our PM2D code is capable of simulating LPIs at ICF experimental scales.

## 2、Physical model

The physical model of our PM2D code is based upon the classical theory of laser plasma interactions [35]. The wave equations to describe the propagation of the laser beams in plasma are given by

$$(1) \quad \begin{aligned} \frac{\partial^2 a_x}{\partial t^2} - \nabla^2 a_x &= -4\pi^2 n_e a_x, \\ \frac{\partial^2 a_y}{\partial t^2} - \nabla^2 a_y &= -4\pi^2 n_e a_y, \\ \frac{\partial^2 a_z}{\partial t^2} - \nabla^2 a_z &= -4\pi^2 n_e a_z, \end{aligned}$$

where $\nabla^2 = \partial^2/\partial x^2 + \partial^2/\partial y^2$, and the plasma motion is limited in the $x - y$ plane, $a_x, a_y, a_z$ are the $x, y, z$ components of the vector potential, respectively, that are normalized as $\boldsymbol{a} = e\mathbf{A}/(m_e c^2)$, where $\mathbf{A}$ is the vector potential. $n_e$ is electron density normalized to $n_c$, where $n_c = \omega_0^2 m_e/(4\pi e^2)$ is the critical density corresponding to the incident laser frequency of $\omega_0$. The space and time are normalized to $2\pi c/\omega_0$ and $2\pi/\omega_0$, respectively. In the two dimensional geometry, the polarization planes of incident laser beams can be divided into P polarization and S polarization that are parallel and orthogonal with the simulation plane ($x - y$ plane), respectively. In the P polarization condition, when laser beams are injected with a nonzero angle with respect to the default propagation direction of $x$, there will be a component of vector potential in the $x$ direction. Therefore, to simulate the P and S polarizations as well as the oblique incidence of laser beams, the time evolutions of the distributions of three vector potential components $a_x$, $a_y$ and $a_z$ in the x-y plane should be all considered as shown in Fig. 1(a).

In Eq. (1), the vector potential of $\boldsymbol{a}$ is a superposed value including both incident and scattering laser contributions, namely, $\boldsymbol{a} = \boldsymbol{a}_s + \boldsymbol{a}_i$, here $\boldsymbol{a}_i$ denotes the vector potential of the incident laser and $\boldsymbol{a}_s$ denotes the vector potential of the scattering light. The right terms in the wave equations act as the wave sources, namely, when the plasma waves (fluctuations of $n_e$) are excited, the incident laser light will couple with the plasma waves to produce the scattering light, and the scattering light can also couple with the plasma wave to dissipate the incident laser. Treating the incident and scattering laser lights together by a single wave equation enables us to capture the secondary scattering and anti-Stokes scattering of laser lights. In the conventional three wave model, however, the incident laser lights and scattering laser lights are usually treated by two individual wave equations, which restricts the physical model to be applicable only to the simple primary LPI process. In contrast, we describe the incident laser light and scattering laser light together in a single wave equation, which makes it convenient to simulate not only the simple LPI processes, but also the nonlinear phenomena such as the formation of density cavities, filamentation of laser beams and so on.

It should be pointed out that the distinct difference between our PM2D code and conventional PIC codes is that the driven forces of macro particles in the PM2D code are fluid forces obtained from the fluid model of LPIs, while the driven force in the PIC code is the Lorentz force [38]. The motion equations of macro particles in the interaction of the plasma with lasers are described by [35]

$$(2) \quad \begin{aligned} \frac{d\mathbf{u}_e}{dt} &= 4\pi^2 \mathbf{E} - \frac{1}{2}\nabla(a_x^2 + a_y^2 + a_z^2) - \frac{2v_{eth}^2}{n_e}\nabla n_e, \\ \frac{d\mathbf{u}_i}{dt} &= -4\pi^2 \frac{m_e}{m_i}\mathbf{E} - \frac{2v_{ifh}^2}{n_i}\nabla n_i, \end{aligned}$$

where $\mathbf{u}_e = (u_{ex}, u_{ey})$ is the velocity vector of electrons that is normalized to the speed of light $c$ in the vacuum. $\mathbf{E} = (E_x, E_y)$ is the electrostatic field that is normalized to $2\pi c\omega_0 m_e/e$, and $v_{eth} = \sqrt{T_e/511}$, $v_{ith} = \sqrt{T_i/511/m_i}$ are the thermal velocity of electron and ion normalized to $c$, $T_e$ and $T_i$ are normalized to 1keV. $n_i$ is the ion density normalized to the critical density $n_c$, $m_e$ and $m_i$ are electron and ion mass normalized to the electron mass. As we solve the momentum equations by the particle-mesh method, the accelerations of macro particles are required, which can be given by the Lagrangian time derivative of $d\mathbf{u}_{e,i}/dt = \partial \mathbf{u}_{e,i}/\partial t + \nabla(\mathbf{u}_{e,i} \cdot \mathbf{u}_{e,i})$, here the subscript of $e$ and $i$ denotes electrons and ions, respectively. In the particle-mesh method, the plasma density can be updated according to the instantaneous distribution of macro particles and it is unnecessary to solve the continuity equation.

As mentioned above, $\mathbf{a} = \mathbf{a}_s + \mathbf{a}_i$ is the superposed light field that includes both the incident and scattering laser lights, therefore the ponderomotive force $\nabla(a^2)$ in Eq. (2) can be expressed as $\nabla(a_s^2 + a_i^2 + 2\mathbf{a}_s\mathbf{a}_i)$. Here the terms $\nabla a_i^2$ and $\nabla a_s^2$ give the pondermotive forces induced by nonuniform spatial-temporal distributions of the incident and scattering laser intensities, respectively, they are responsible for the formation of density cavity and filamentation of laser beams. The term $2\nabla \mathbf{a}_s\mathbf{a}_i$ gives the pondermotive force induced by the beat wave of incident and scattering lights that is responsible for the excitation of LPIs. Including the filamentation and self-focusing effects of laser beams in the LPI model is important, since it has been already found that the self-focusing of incident laser beams may remarkably increase the local laser intensity and result in a strong LPI excitation in the laser self-focusing area [36]. Moreover, the formation of a density cavity around the quarter critical density has been found to play an important role in the evolution of TPD and SRS [37].

Finally, the Poisson equation that describes the spatial distribution of the electrostatic potential of $\varphi$ is adopted to close the governing equations, that is

$$(3) \quad \nabla^2 \varphi = n_i - n_e,$$

where $\varphi$ is normalized to $\omega_0^2 m_e/e$. After calculating the electrostatic potential of $\varphi$, the electrostatic field of $\mathbf{E}$ can be obtained by taking the divergence of $\varphi$, namely $\mathbf{E} = -\nabla\varphi$.

Equations (1) - (3) form the physical model of our PM2D code, which is able to describe several kinds of parametric instabilities like SRS, SBS, CBET as well as some other instabilities like the laser filamentation instability in laser-plasma interactions. Moreover, our physical model can capture the nonlinear effects of LPI such as the secondary scattering, harmonic generation, frequency shift, nonlinear saturation, etc. [12], which are hard to be captured by the three-wave model. More importantly, since the momentum equations of electrons and ions are solved by the particle-mesh method, the kinetic effects of LPIs can be included self-consistently in our PM2D code, while the phenomenological methods are usually adopted to treat the kinetic effects in the general fluid models. Comparing with the conventional PIC codes, the PM2D code is based on a specifically constructed physical model for the LPIs, and hence the required macro-particles per grid can be greatly decreased in the PM2D code [39]. As a result, the simulation cost of PM2D code is much smaller than that of the PIC codes.

## 3 Numerical methods

### 3.1 Laser module

In the PM2D simulation, a plasma slab is set at the center of the simulation box, and the vacuum space is usually set around the plasma slab for the convenience of diagnosing the scattering laser light and allowing the plasma expanding in the evolution of LPIs. The laser beams are injected from the left boundary of the simulation box and propagate in the $x-y$ plane with their own wave vectors **k**, as shown in Fig. 1(a). The normalized vector potential **a** of a laser beam lies in the vertical plane of its wave vector **k**. For the oblique incidence and arbitrary polarization, the vector potential **a** of an incident laser beam may have components in all of three directions of $x, y$ and $z$ as shown in Fig. 1(a). Assuming the direction of vector potential **a** has an angle of $\theta$ with respect to the $z$-axis, then the components of **a** in the S-polarization direction ($z$ direction) is $a_{sz} = a\cos(\theta)$, and in the P-polarization plane ($x-y$ plane) is $a_p = a\sin(\theta)$. if $a_p$ has a angle of $\beta$ with respect to the $x$ direction in the P polarization plane, the components of the vector potential in $x$ and $y$ directions can then be expressed as $a_{px} = a\sin(\theta)\cos(\beta)$ and $a_{py} = a\sin(\theta)\sin(\beta)$, respectively, as shown in Fig. 1(a).

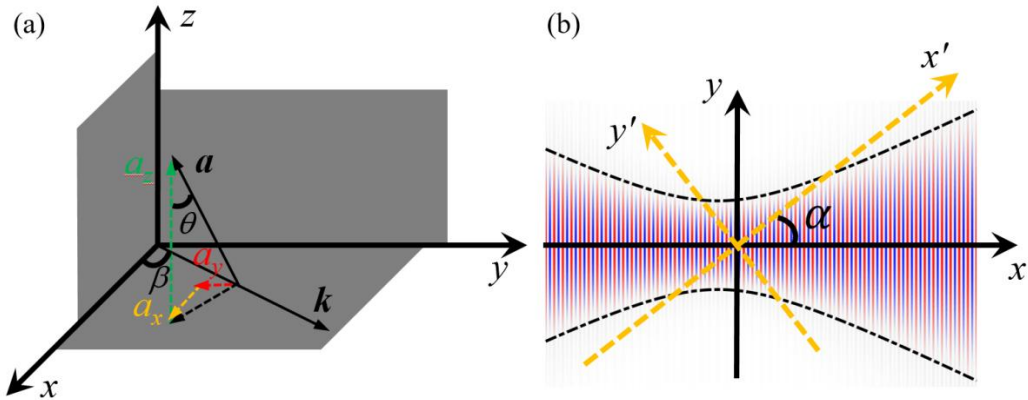

Figure 1: (a) The vector relationship among the normalized vector potential $a$ and its components $a_x$, $a_y$, $a_z$. Here $k$ is the wave number that points to the laser propagation direction, θ is the angle between $a$ and z-direction and β is the angle between $k$ and x-direction. The vector potential $a$ may point to any directions within the vertical plane of $k$ and thus its components in the x,y and z directions should all be simulated. (b) A schematic diagram that shows how to load a laser beam on the simulation boundary to consider the oblique incidence. Assuming $x' - y'$ is the coordinate system of the simulation domain, then the Gaussian laser beam are obliquely injected in this coordinate system with an angle of α with respect to the $x'$ axis. The boundary condition of the obliquely incident laser beam on the $y'$ axis can be obtained by firstly transforming the coordinates of the $y'$ axis to the coordinates (x,y) and then substituting (x,y) into the expression of Gaussian laser beam (Eq.(4)) in the $x - y$ coordinate system.

In the PM2D code, the laser beam is injected on the left boundary of the simulation box, both of plane wave and Gaussian beam can be assumed. The expression of a Gaussian beam can be written as

$$(4) \quad a = \exp[-(y-y_0)^2/w_0^2]\cos(\omega t - kx),$$

where $y_0$ is the beam center and $w_0$ is the characteristic beam width of a Gaussian laser beam. For simplicity, the focal plane is also set on the $y$ axis. For the normal incident condition, the boundary condition on the $y$-axis at the $(n+1)$-th time step can be obtained by directly substituting $x = 0$ and $t = (n+1)dt$ in Eq. (4). For the oblique incidence of a laser beam with an angle of $\alpha$ with respect to the $x$ direction, we first introduce a set of new coordinate system of $x' - y'$, which can be obtained by counter-clockwise rotation of the $x - y$ coordinate system in an angle of $\alpha$, as shown in Fig. 1(b). By doing so, the normal incident laser beam in the $x - y$ coordinate system is oblique incident in the $x' - y'$ system. To obtain the boundary condition (vector potential values) of an oblique incident laser beam on the $y'$ axis, the coordinates $(x' = 0, y')$ on the $y'$ axis should be firstly converted to the coordinates in the $x - y$ system by a coordinate transformation method:

$$(5) \quad \begin{array}{l} x = x'\cos(\alpha) - y'\sin(\alpha) \\ y = x'\sin(\alpha) + y'\cos(\alpha) \end{array},$$

where $(x' = 0, y')$ are the coordinates of the $y'$ axis in the $x' - y'$ coordinate system, and $(x, y)$ are the corresponding coordinates in the $x - y$ coordinate system. The obtained $(x, y)$ can then be substituted into Eq. (4) to get the boundary condition of an oblique incidence laser beam.

In the interior region of the simulation box, the wave equation is discretized by the second-order central difference scheme for both time derivative and spatial derivative, namely, Eq. (1) is discretized as [40]:

$$(6) \quad \frac{a_{i,j}^{n+1} - 2a_{i,j}^n + a_{i,j}^{n-1}}{dt^2} - \frac{a_{i+1,j}^n - 2a_{i,j}^n + a_{i-1,j}^n}{dx^2} - \frac{a_{i,j+1}^n - 2a_{i,j}^n + a_{i,j-1}^n}{dy^2} = -4\pi^2 n_{ei,j}^n a_{i,j}^n,$$

where $a$ denotes either $a_x, a_y$ or $a_z$. The subscript of $i$ and $j$ denotes the $i$-th and $j$-th grid in $x$ and $y$ direction respectively, the superscript of $n$ denotes the $n$-th time step, $dt$ is the time step, and $dx, dy$ are the grid sizes in the $x$ and $y$ directions, respectively.

In the LPI process, the scattering laser lights will propagate away from the fusion target. To simulate the escaping of the scattering lights, the laser lights that arrive at the boundaries of the simulation domain should be absorbed rather than be reflected back into the simulation domain. Considering that the scattering lights may distribute within a wide range of angles, the absorbing boundary condition should be employed on all of four boundaries. In our PM2D code, a second-order absorbing boundary condition is adopted [41]. Here we set the left boundary as an example to explain the setting of the absorbing boundary condition. The partial differential equation that describes the absorption of laser light on the left boundary can be written as [41]

$$(7) \quad \frac{\partial^2 a}{\partial x \partial t} - \frac{\partial^2 a}{\partial t^2} + \frac{\partial^2 a}{\partial y^2}\bigg|_{x=0} = 0,$$

where $a_{x,y,z}$ denotes the x, y, or z component of the vector potential and this equation is only fulfilled on the left boundary of $x = 0$. Equation (7) can be discretized as [41]:

$$(8) \quad \left(\frac{a_{1,j}^{n+1} - a_{1,j}^{n-1}}{2dtdx} - \frac{a_{0,j}^{n+1} - a_{0,j}^{n-1}}{2dtdx}\right) - \frac{1}{2}\left(\frac{a_{0,j}^{n+1} - 2a_{0,j}^n + a_{0,j}^{n-1}}{dt^2} + \frac{a_{1,j}^{n+1} - 2a_{1,j}^n + a_{1,j}^{n-1}}{dt^2}\right) + \frac{1}{4}\left(\frac{a_{0,j+1}^n - 2a_{0,j}^n + a_{0,j-1}^n}{dy^2} + \frac{a_{1,j+1}^n - 2a_{1,j}^n + a_{1,j-1}^n}{dy^2}\right) = 0,$$

where $a_{i,j}^n$ denotes $a(idx, jdy, ndt)$, and the subscript of 0 and 1 denotes the boundary grids of $i = 0$ and the adjacent interior grids of $i = 1$, respectively. Noting that in the calculation of the boundary condition on the points of $i = 0$ at the $(n + 1)$-th time step by using Eq. (8), the vector potential values on the grids $i = 1$ at the $(n + 1)$-th time step are required. Therefore, the values of vector potential on the interior grids of $i = 1$ at the $(n + 1)$-th time step should firstly be calculated by Eq. (6) using the boundary conditions at the $n$-th time step. Then the boundary condition on the points of $i = 0$ at the $(n + 1)$-th time step can be updated by using Eq. (8). The stability condition for the discrete scheme of the second-order absorbing boundary condition is $dt/dx < \sqrt{2}/2$, which is in line with the stability condition of the discrete scheme of Eq. (6).

From the derivation of the absorbing boundary condition of Eq. (8), one can find that the vector potential **a** in Eq. (8) should actually be the left traveling waves only. However, when the laser beams are loaded on the left boundary, the vector potential on the interior points of i = 1 may include both left-traveling waves and right-traveling waves due to **a** is a supposed field that includes both the incident and scattering light contributions. Therefore, to implement the absorbing boundary condition correctly at (n + 1)-th time step using Eq. (8), the left traveling wave in the superposed laser field on the grids of i = 1 should be isolated firstly. In our PM2D code, the right traveling wave around the left boundary is only the incident laser light. Noting that the grids of $i = 1$ lie in the vacuum region, the right traveling waves on the grids of $i = 1$ at the $(n + 1)$-th time step can be obtained analytically from the vector potential formulas of the incident laser beams. A similar calculation procedure can be followed as the simulation of the boundary values for the oblique incidence laser beams using Eq. (6). Following the above procedure, the left boundary condition that simultaneously guarantees the free injection of pump laser light and the sufficient absorption of scattering lights has been constructed, namely, $a_{0,j}^{n+1} = a_{L0,j}^{n+1} + a_{R0,j}^{n+1}$, where $a_{L0,j}^{n+1}$ and $a_{R0,j}^{n+1}$ denote the left traveling waves (scattering laser light) and right traveling waves (incident laser light) on the left boundary at the $(n + 1)$-th time steps, respectively.

## 3.2 Electrostatic module

The electrostatic field is crucial in the numerical simulation of LPIs since it not only determines the oscillating frequency of plasma waves but also gives birth to nonlinear and kinetic behaviors of plasma waves. Therefore, the electrostatic field should be calculated with high accuracy. The potential of the electrostatic field is given by Poisson's equation, as shown in Eq. (3). In our study, the Poisson equation is solved by a Fourier transformation method along the $y$ direction to decrease the simulation cost. Assuming the density distributions of electrons and ions are periodic in the $y$ direction with a cycle length of $L = y_{max} - y_{min}$, the electrostatic potential is also a periodic function that can be expanded by the Fourier series in the $y$ direction as:

$$(9) \quad \varphi(x, y) = \sum_{l=-\infty}^{l=\infty} \hat{\varphi}(x, l) e^{j\left(\frac{2\pi l}{L}\right)y}, y \in [y_{min}, y_{max}],$$

where $j$ is the imaginary unit, $\hat{\varphi}(x, l)$ is the Fourier coefficient, $l$ is the order of Fourier series, and $y_{max}$ and $y_{min}$ are the y coordinates of the upper and lower boundaries, respectively. Substituting Eq. (9) into Eq. (3), one can get a differential equation system for different $l$ (or $k_y$), that is:

$$(10) \quad \frac{\partial^2 \hat{\varphi}}{\partial x^2} - k_y^2 \hat{\varphi} = \hat{n}_i - \hat{n}_e, k_y = \frac{2\pi l}{L}, l = -\frac{N}{2}...0...\frac{N}{2},$$

here $k_y$ is the wave number in the $y$ direction, $\hat{n}_e$ and $\hat{n}_i$ are the Fourier decompositions of the electron and ion densities, respectively, and $N = (y_{max} - y_{min})/dy$ is the total grid points in the $y$ direction. The series of differential Eqs. (10) are independent of each other, so it is convenient to solve them in parallel. In Eqs. (10), the Fourier coefficients of $\hat{n}_e$ and $\hat{n}_i$ can be obtained using the fast Fourier transform method. After $\hat{n}_e$ and $\hat{n}_i$ are obtained, Eqs. (10) can then be discretized in the $x$-direction by a central difference scheme for different wave numbers $k_y$, namely:

$$(11) \quad \frac{\hat{\varphi}_{i+1} - 2\hat{\varphi}_i + \hat{\varphi}_{i-1}}{dx^2} - k_y^2 \hat{\varphi}_i = (\hat{n}_i - \hat{n}_e)_i, k_y = \frac{2\pi l}{L}.$$

For each $k_y$ or $l$, the coefficient matrix of the linear equation system of $\hat{\varphi}_i$ in Eq. (11) forms a tridiagonal matrix that can be solved efficiently by a chasing method. After the electrostatic potential in the spectrum space is obtained through Eqs. (11), the electrostatic potential in the $x - y$ space can be calculated by an inverse Fourier transform method. The electrostatic field can then be generated by taking the divergence of electrostatic potential, namely, $E_x = \partial\varphi/\partial x, E_y = \partial\varphi/\partial y$. To mitigate the numerical noise in the calculation of the electrostatic field, $E_x$ and $E_y$ are set on the center of the grids of the electrostatic potential. Under this setting, $E_x$ is first calculated in the $x$ direction by a central difference scheme and then averaged in the $y$ direction, namely:

$$E_{xi+1/2,j+1/2} = \frac{1}{2}\left(\frac{\varphi_{i+1,j+1} - \varphi_{i,j+1}}{dx} + \frac{\varphi_{i+1,j} - \varphi_{i,j}}{dx}\right).$$

Similarly, $E_y$ can be calculated as:

$$E_{yi+1/2,j+1/2} = \frac{1}{2}\left(\frac{\varphi_{i+1,j+1} - \varphi_{i+1,j}}{dy} + \frac{\varphi_{i,j+1} - \varphi_{i,j}}{dy}\right),$$

where the subscript $i + 1/2$ and $j + 1/2$ denote the central points of the grid cell of the electrostatic potential.

## 3.3 Particle module

In general, the conventional fluid models based on the fixed Eulerian grids are hard to capture the kinetic effects of LPIs, such as the generation of hot electrons that in return plays an important role in the evolution of LPIs. To overcome the disadvantages of the fluid description, the particle-mesh (PM) method [42] is adopted to simulate the plasma dynamics in our studies. In the PM method, the macro particles can move freely across the fixed grids, which makes it convenient to describe the kinetic effects such as the particle trapping and hot electron generation [43].

In the PM2D code, the electron and ion macro-particles are initially distributed evenly within the grids. Assuming there are $N$ and $M$ macro-particles per grid in $x$ and $y$ directions respectively, then the spatial intervals between two adjacent macro-particles in $x$ and $y$ directions are $dx/N$ and $dy/M$, respectively. Correspondingly, the density weight of a macro-particle can be calculated by:

$$n_{sp}^0 = \frac{n_s^0(idx, jdy)}{NM}, s = e, i,$$

where $n_{sp}$ is the density weight of the $p$-th macro-particle that lies in the grid cell of $x = idx, y = jdy$ and $s$ denotes the species of electrons or ions. The superscript of 0 denotes the 0-th time step and $n_s^0(idx, jdy)$ is the density for the $s$ species on the $i$-th and $j$-th grid point. The drift velocity of macro-particles can also be given according to the initial velocity distribution of electron and ion on the grid, that is:

$$\mathbf{v}_{sp}^0 = \mathbf{v}_s^0(i,j), s = e, i,$$

where $\mathbf{v}_{sp}$ denotes the velocity of the $p$-th macro-particle and $\mathbf{v}_s^0(i,j)$ is the initial drift velocity on the $i$-th and $j$-th grid. From the above procedure, the position, density weight, and drift velocity of each macro-particles are initialized. In the development of LPIs, the velocities and positions of macro-particles change with time due to the driving of fluid forces. However, the density weight of every macro-particle keeps unchanged all the time. The distributions of plasma density and fluid velocity on the grids can be obtained by depositing particle densities and velocities on the grids respectively by an area weighting interpolation method. Conversely, the drive forces on the grids can be interpolated to particle according to the particle positions.

### 3.3.1 Interpolation method between grid values and particle values

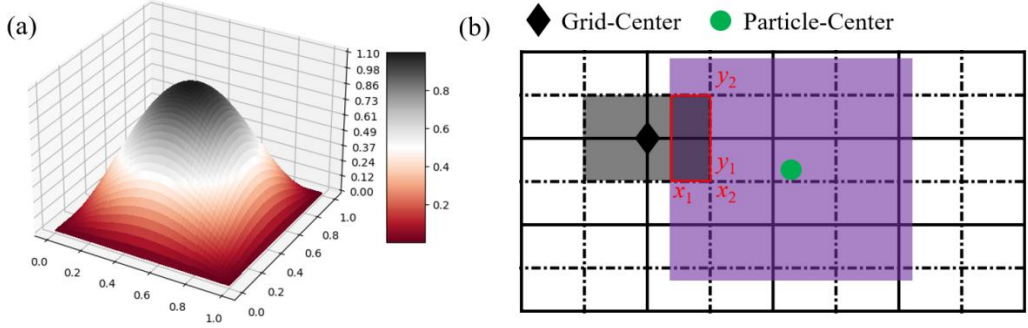

Figure 2: (a) The shape function of the macro-particles used in the PM2D code, which is constructed by a multiplication of two sine functions in x and y directions as defined by Eq. (12). (b) A schematic diagram that shows the area weighting interpolation method in the PM2D code, where the black shaded area denotes the control area of a grid point and the purple shaded area denotes the influence region of a macro-particle. The interpolation between a grid point and a macro-particle is carried out according to the four corner points of the overlapped region

In order to decrease the computational cost, usually only a small number of macro-particles per grid are used in the PM2D code [38]. Correspondingly, a high-order shape function of the macro-particles is adopted to control the numerical noise to a low level in this case:

In order to control the numerical noise to a low level even when a small number of macro-particles per grid is used to decrease the computational cost [38], a high-order shape function is adopted for the macro-particles in the PM2D code, namely:

$$(12) \quad S(x,y) = \left(\frac{D}{\pi}\right)^2 \sin\left(\frac{\pi}{2D}x\right) \sin\left(\frac{\pi}{2D}y\right), x, y \in [0, 2D],$$

where $S(x, y)$ is the shape function that defined on a square area with a boundary length of 2D as shown in Fig. 2(a). For convenience, $S(x, y)$ is chosen as an integrable function so that one can analytically calculate the volume weighting on the overlapping area of particle and grid, namely:

$$(13) \quad w_p = \int_{y_1}^{y_2} \int_{x_1}^{x_2} \left(\frac{D}{\pi}\right)^2 \sin\left(\frac{\pi}{2D}x\right) \sin\left(\frac{\pi}{2D}y\right) = \frac{1}{4}\left[\cos\left(\frac{\pi}{2D}x_1\right) - \cos\left(\frac{\pi}{2D}x_2\right)\right]\left[\cos\left(\frac{\pi}{2D}y_1\right) - \cos\left(\frac{\pi}{2D}y_2\right)\right],$$

here $w_p$ is the volume weighting of a rectangular overlap region, whose minimum coordinates and maximum coordinates in x and y directions are $x_1, y_1$ and $x_2, y_2$, respectively, as shown in Fig. 2(b). In Eq. (13), $w_p$ on the whole particle area ($x, y \in [0, 2D]$) is normalized to 1 to conveniently calculate the volume weighting on the overlaping region. In the area weighting interpolation method, we set a rectangular control volume for every grid point. The control volume has a length of $dx/2$ and $dy/2$ in each sides of the center grid point in x and y directions, respectively. The physical values within this control volume are assumed to be same as the center grid point.

In the deposition process from macro-particle values to grid values, for each grid point, one needs to consider the contributions from all related macro-particles that have overlapping regions with this grid point. Assuming on a grid point, there are $q$ related macro-particles, then the physical value (here we take electron density as an example) on this grid point can be calculated as:

$$(14) \quad n_{ei,j} = \frac{1}{4}\sum_{p=1}^{p=q}\left[\cos\left(\frac{\pi}{2D}x_{1p}\right) - \cos\left(\frac{\pi}{2D}x_{2p}\right)\right]\left[\cos\left(\frac{\pi}{2D}y_{1p}\right) - \cos\left(\frac{\pi}{2D}y_{2p}\right)\right]n_{ep},$$

where $n_{eij}$ is the electron density on the $i$-th and $j$-th grid point, and $n_{ep}$ is the density weight of the $p$-th electron particle. $x_{1p}, x_{2p}, y_{1p}, y_{2p}$ give the corner points of the rectangular overlapping region as shown in Fig. 2(b). Generally, it is inefficient to take the grid points as the loop variable to calculate the density on the mesh grids by using Eq. (14). In our PM2D code, we set macro particles as the loop variable, namely, every particle will deposit its density on the mesh grids and their sum gives the whole density distribution.

In the interpolation process from the fluid forces on the grid points to a macro particle, one needs firstly find the possible grid points whose control volumes have overlapping regions with the chosen macro particle. Assuming the grid points $(x_{i'}, y_{j'})$ with the subscripts changing from $i' = s1, j' = g1$ to $i' = s2, j' = g2$ overlap with the chosen macro-particle, then the total force acting on the chosen macro-particle can be calculated as:

$$(15)\ \mathbf{F}_p = \frac{1}{4}\sum_{i'=s1,j'=g1}^{i'=s2,j'=g2}\left[\cos\left(\frac{\pi}{2D}x_{1i',j'}\right) - \cos\left(\frac{\pi}{2D}x_{2i',j'}\right)\right]\left[\cos\left(\frac{\pi}{2D}y_{1i',j'}\right) - \cos\left(\frac{\pi}{2D}y_{2i',j'}\right)\right]\mathbf{F}_{i',j'},$$

where $\mathbf{F}_p$ denotes the total drive force on the chosen macro-particle, $i'$, $j'$ denotes the $i'$-th and $j'$-th grid point that has overlapping region with the chosen macro-particle, $\mathbf{F}_{i',j'}$ is the drive force on the mesh grid $x = i'dx$, $y = j'dy$, and $x_{1i',j'}, x_{2i',j'}, y_{1i',j'}, y_{2i',j'}$ also gives corner points of the rectangular overlapping region.

In the PM2D code, we adopted the second order Runge-Kutta method to calculate the positions and velocities of macro-particles at the $(n + 1)$-th time step [44]. To this end, at the $n$-th time step, we firstly adopt the interpolating method in Eq. (14) to deposit the densities and momentums of all macro particles to the mesh grids. As a result, the density and velocity distributions on grids can be obtained. Combing with the vector potentials of laser lights at the $n$-th time step, the fluid forces on the mesh grids can be calculated according to the momentum equation in Eq. (2), namely:

$$(16) \quad \begin{aligned}\mathbf{F}_e^n &= 4\pi^2\mathbf{E}^n - \frac{1}{2}\nabla\left(a_x^2 + a_y^2 + a_z^2\right)^n - \frac{2v_{eth}^2}{n_e^n}\nabla n_e^n,\\ \mathbf{F}_i^n &= -4\pi^2\frac{m_e}{m_i}\mathbf{E}^n - \frac{2v_{ith}^2}{n_i^n}\nabla n_i^n,\end{aligned}$$

where the superscript of $n$ denotes the $n$-th time step, and the divergence terms are discretized by a central difference scheme. The fluid forces on the grid points are then interpolated to every macro-particles by using the interpolating method in Eq. (15). The velocity and position of each macro-particle at the $(n + 1)$-th time step can be preliminarily calculated as:

$$\begin{aligned}
\widehat{\mathbf{u}}_{ep}^{n+1} &= \mathbf{u}_{ep}^n + dt\mathbf{F}_{ep}^n, \\
\widehat{\mathbf{x}}_{ep}^{n+1} &= \mathbf{x}_{ep}^n + \frac{dt}{2}\left(\widehat{\mathbf{u}}_{ep}^{n+1} + \mathbf{u}_{ep}^n\right).
\end{aligned} \quad (17)$$

Here we select the electron macro-particle as an example. In Eq. (17), $\mathbf{F}_{ep}^n$ is the total driven force that acts on the $p$-th macro particle, and $\widehat{\mathbf{u}}_{ep}^{n+1}, \widehat{\mathbf{x}}_{ep}^{n+1}$ denotes the velocity and position of $p$-th particle at the $(n+1)$-th time step, respectively. It is worth pointing out that $\widehat{\mathbf{u}}_{ep}^{n+1}$ and $\widehat{\mathbf{x}}_{ep}^{n+1}$ obtained from Eq. (17) only have first order accuracy in time. Substituting $\widehat{\mathbf{u}}_{ep}^{n+1}$ and $\widehat{\mathbf{x}}_{ep}^{n+1}$ into Eq. (14), the density and velocity on the grids at the $(n+1)$-th time step can be obtained. Combining with the vector potentials of laser lights at the $(n+1)$-th time step, then the fluid forces on the grids at the $(n+1)$-th time step can be calculated as:

$$\begin{aligned}
\mathbf{F}_e^{n+1} &= 4\pi^2 \mathbf{E}^{n+1} - \frac{1}{2}\nabla\left(a_x^2 + a_y^2 + a_z^2\right)^{n+1} - \frac{2v_{eth}^2}{n_e^{n+1}}\nabla n_e^{n+1} \\
\mathbf{F}_i^{n+1} &= -4\pi^2 \frac{m_e}{m_i}\mathbf{E}^{n+1} - \frac{2v_{ith}^2}{n_i^{n+1}}\nabla n_i^{n+1}
\end{aligned} \quad (18)$$

After the total drive force on the grids is obtained by Eq. (18), the force that acts on every macro particles at the $(n+1)$-th time step can be obtained by the interpolation method using Eq. (15). Finally, the velocities and positions of macro particles that have second-order accuracy can be obtained as:

$$\begin{aligned}
\mathbf{u}_{ep}^{n+1} &= \frac{1}{2}\left(\widehat{\mathbf{u}}_{ep}^{n+1} + \mathbf{u}_{ep}^n\right) + \frac{1}{2}dt\mathbf{F}_{ep}^{n+1} \\
\mathbf{x}_{ep}^{n+1} &= \mathbf{x}_{ep}^n + \frac{1}{2}dt\left(\mathbf{u}_{ep}^{n+1} + \mathbf{u}_{ep}^n\right)
\end{aligned} \quad (19)$$

From Eqs. (16)-(19), the physical quantities on the macro particles and grid points can be updated successively. In general, the process of updating the velocities and positions of macro-particles is the most time-consuming process in the PM2D code. To decrease the computational cost, the second order Runge-Kutta method is adopted which allows a time step several times larger than that used in the wave equations.

## 4 Program Flow

The flow chart of the PM2D code is illustrated in Fig. 3.

(1) The distribution of plasma density and velocity on the grids at the $n$-th time step are calculated by depositing macro-particle densities and velocities onto the grids by using Eq. \eqref{to_grids}.

(2) The electron density on the grids at the $n$-th time step is then substituted into the wave equation to obtain the vector potential of total light field at the $(n+1)$-th time step by using the second-order absorbing boundary condition. Simultaneously, the electrostatic fields on the mesh grids at the $n$-th time step can be updated according to the charge separation between the electrons and ions at the $n$-th time step, as shown in Sec.\,\ref{3.2}.

(3) Using the vector potential, electrostatic field and plasma density at the $n$-th time step, the fluid force on the mesh grids at $n$-th time step can be calculated by Eq. \eqref{momentum equation}, which is subsequently used to calculate the total driven force acting on each macro particle by the interpolation method using Eq. \eqref{to_particle}.

(4) The intermediate velocities $\hat{\mathbf{u}}_{ep}^{n+1}$ and positions $\hat{\mathbf{x}}_{ep}^{n+1}$ of every macro particles at the $(n+1)$-th time step are calculated with the first-order accuracy by using \eqref{first_rung}.

(5) The plasma density and velocity distributions on the mesh grids at the $(n+1)$-th time step are updated by using $\hat{\mathbf{u}}_{ep}^{n+1}$ and $\hat{\mathbf{x}}_{ep}^{n+1}$ according to \eqref{to_grids}.

(6) The newly obtained plasma density and velocity at the $(n+1)$-th time step are then used to calculate the electrostatic field on grid at the $(n+1)$-th time step.

(7) Using the vector potential, electrostatic field, and plasma density at the $(n+1)$-th time step, the drive force on the mesh grids at the $(n+1)$-th time step can be obtained, which can be used again to get the total driven force upon every macro-particles by the interpolation method using Eq. \eqref{to_particle}.

(8) Finally, the velocities and positions of macro particles at the $(n+1)$-th time step can be obtained by using Eq. \eqref{velocities} with the second-order accuracy.

(9) The above loop is continuously executed until the final simulation time comes.

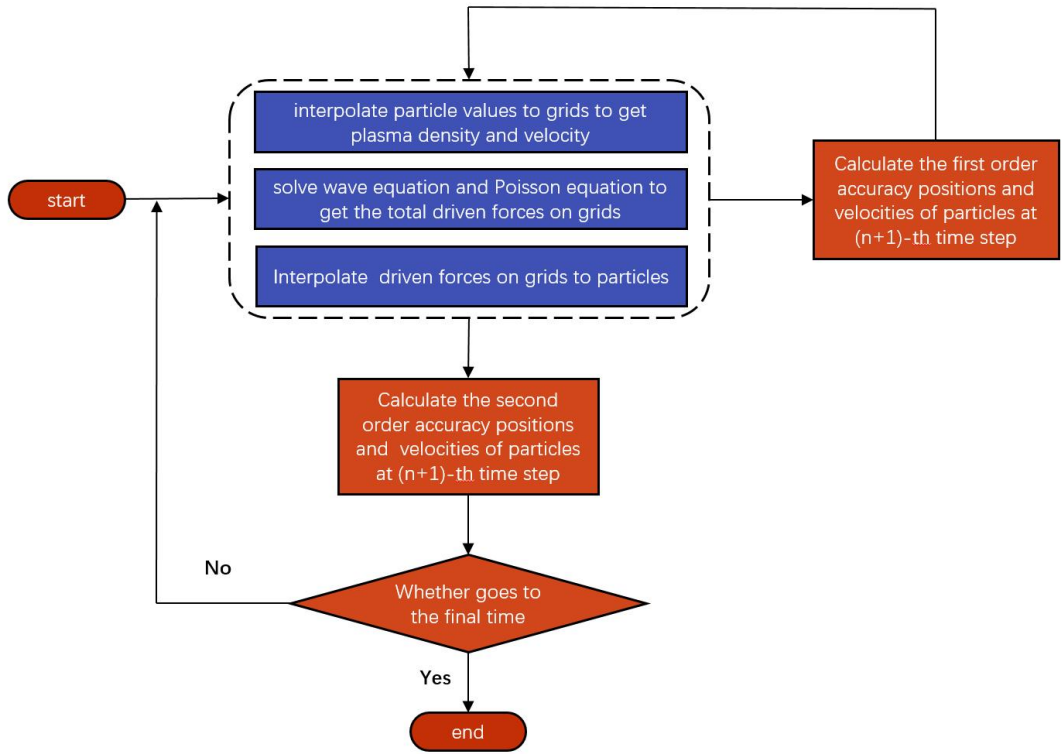

Figure 3: The flow chart of the PM2D code, where the module in the black dashed box is called for two times in a time step.

# 5 GPU-based parallelization

## 5.1 Test platform

As a typical CPU, a GPU also has its Arithmetic logic unit, control unit, cache, and DRAM. A modern NVIDIA GPU contains tens of multiprocessors (MPs). A multiprocessor consists of 8 scalar processors (SPs), each capable of executing an independent thread. Threads are grouped together as a grid of thread blocks, such that each MP executes one or more thread blocks and each of its SP runs one or more threads within a block. Our experimental platform consists of 16 nodes with a total of 64 NVIDIA A100 graphics cards. Each graphics card is based on the NVIDIA CUDA model, which is also the name of the software for programming under this architecture. All of 64 A100 GPUs contain in total 6912 cores organized in 108 streaming multiprocessors . It provides 40 GB of device memory with a memory bandwidth of 1555 GB/s, accessible by all its cores and the CPU through the Kunpeng 920 with 96 cores.

## 5.2 Cuda device

Groups of 32 threads (warps) within a block are scheduled with the same instruction in a single cycle of clock time. Furthermore, threads within a block can cooperate by sharing data through some shared memory and must synchronize their execution to coordinate the memory accesses. In contrast, there is no synchronization between thread blocks in a grid of blocks. Hence, the kernels only work with the data in the GPU device's memory, and their final results must be transmitted to the CPU. However, the main bottleneck in the CUDA architecture lies in the data transfer between the host (CPU) and the device (GPU). Any transfer between CPU and GPU may reduce the time execution performance. To summarize, it is necessary to minimize the communication between the CPU and GPU.

## 5.3 Code parallelization

Our study employs the domain decomposition method to partition a 2D grid with $Nx \times Ny$ dimensions into $m$ parts along the x-coordinate direction. Each part represents a sub-domain; the total number of sub-domains is denoted as $m$. Each sub-domain is assigned to a corresponding process equipped with a GPU for computation. Two principles guide the domain decomposition strategy. Firstly, $m$ equals the total number of GPUs in use. Secondly, the shape of each sub-domain should be as square-like as possible to reduce the amount of information exchanged between GPUs. Throughout the simulation, the flow variables of each sub-domain persistently reside in the global memory of the assigned GPU. From the beginning to the end of the simulation, the computed flow variables are stored and exported to different files as required.

For the wave equation, we utilize the central difference method in our approach to advancing the time step. Each GPU thread is responsible for reading and updating a specific grid node. This allows for parallel processing and efficient computation.

For Poisson's equation, we employ a three-step process to solve the modified linear system:

1. Fast Fourier Transform (FFT): Firstly, the data is transformed using the fast Fourier transform. This operation is performed only in the $y$ direction, and the Fourier coefficients are computed using the CUFFT library. This step allows for efficient manipulation of the data in the frequency domain.

2. Distributed Arrowhead Decomposition [45]: Next, we utilize a distributed arrowhead decomposition method to decompose the tridiagonal linear system in the $x$ direction. This technique facilitates solving the linear tridiagonal equations for each sub-domain on the corresponding GPU. We leverage the cusparseDgtsv2StridedBatch API from the CUSPARSE library for an efficient and accurate solution of the linear equations.

3. Inverse Fast Fourier Transform (IFFT): Finally, the data is transformed back to the spatial domain using the inverse fast Fourier transform. This operation is also performed only in the $y$ direction, and the Fourier coefficients are calculated once again using the CUFFT library. This step allows us to obtain the final solution of the Poisson equation.

For the particle module, we have divided the computation process into the following main sub-processes:

1. Updating Particle Velocity and Position: This step involves the computation of the velocity and position of each macro-particle. The equations governing the motion of particles are solved to get their updated states.

2. Reconstructing Particle Information within each GPU: After the computation of particle velocities and positions, the particle information within each GPU needs to be reconstructed. This step involves the organization and updating of the data structure that contains the macro-particle for efficient access and further computations.

3. Depositing Electron and Ion Density onto Grid Points: In order to get the electron and ion densities on the grid points, the density contributions of macro-particles are deposited onto the grids. This deposition step is essential for accurately representing the density distribution of particles in the grid-based simulation domain.

For the first subprocess, we have developed a kernel to simulate the motion of macro-particles. In this kernel, each GPU thread is assigned to update the position and velocity of the corresponding macro-particle. For each macro-particle, we find the intersections of their influence area with the grid and save them in shared memory first, then calculate the forces of the macro-particles by the interpolation. After the driving forces for the macro-particles are obtained, the positions and velocities of the macro-particles can then be updated.

After updating the velocities and positions of the macro-particles, it becomes necessary to recount the number of macro-particles within each computation sub-domain and reconstruct the data structure that contains the macro-particles in the second subprocess. This subprocess involves several steps. First, an indicative vector $I$ is created to mark the particles within each process. Particle across the sub-domain boundaries are recorded as 1, while particles remaining within the sub-domain are recorded as 0. Next, a new vector $A$ is generated by performing a scan operation on the indicative vector $I$. This operation aggregates the boundary-crossing information to obtain cumulative counts, which are stored in the vector $A$. Subsequently, a filtering operation is applied to identify particles whose identifiers satisfy the condition $a_{id+1} > a_{id}$ in the vector $a$. The corresponding particle identifiers are marked for further processing. Finally, using the MPI API functions such as $MPI\_SEND$ and $MPI\_RECV$, particles that have crossed the sub-domain boundaries are exchanged with neighboring processes. The received macro-particles are then inserted into the positions previously occupied by the sent macro-particles. Additionally, any extra particles obtained are added at the end of the queue or data structure used to store macro-particles.

For the third subprocess, we have developed a kernel for evaluating the electron and ion density contributions onto the grids. Similar to the first subprocess, in this kernel, we take the macro-particles as the parallel unit. For each macro-particle, we first find the intersections of their influence area with the grid and save them in shared memory, then compute the electron or ion density contributions onto the grids according to the area of the overlapping regions between the macro-particle and the grid volume. To avoid concurrent writing access of the same grid point by multiple threads, we use the atomic function $atomicAdd()$ in CUDA.

In addition, we use the fastest kernels of basic linear algebra subprograms of CUDA (CUBLAS) for vector operations ($cublasDdot, cublasDaxpy$) and apply the $inclusive\_scan$ in the $Thrust$ library to scan the vector $A$. Each kernel inside the main loop of time advancing will be called by the host thread and executed on the GPU platform. Once on the GPU, each kernel will be executed in parallel by multiple threads. In practice, we specify that each block consists of 256 threads in our time-dependence solver. Then, the thread block size in each grid is as follows:

$$Blocks = \frac{N + Threads - 1}{Threads},$$

where $N$ is the number of total threads, and $Threads$ is the size of thread blocks. The size of $N$ has adjusted adaptively for different kernel functions here. Once the time loop execution is finished on the GPU, the cudaDeviceSynchronize function is called for synchronization, and results are copied from the device to the host. The function $cudaFree()$ is used to prevent GPU memory leakage before shutdown.

## 5.4 Simulation speed test and compare with PIC code

*The comparison of the computation time and simulation speed of the PM2D code (GPU code) with those of EPOCH code (CPU code) in the cases using different mesh grids, where the bold numbers show the total computation time in seconds. The ratio of CPU cores to GPUs adopting in all simulations is fixed.*

| Grid number | CPU time (**seconds** × cores) | GPU time (**seconds** × GPU numbers) | CPU time/GPU time |
|---|---|---|---|
| 2000 × 2000 | **1602** × 192 | **67** × 2 | 23.9 |
| 4000 × 4000 | **3087** × 768 | **112** × 8 | 27.5 |
| 6000 × 6000 | **3789** × 1728 | **145** × 18 | 26.1 |
| 8000 × 8000 | **4523** × 3072 | **158** × 32 | 28.6 |
| 10000 × 10000 | **5364** × 4800 | **171** × 50 | 31.8 |

Table 1: The comparison of the computation time of the PM2D code (GPU-based) with that of the EPOCH code (CPU-based) in the cases using different mesh grids, where the bold numbers show the total simulation time in seconds. The ratio of CPU cores to GPUs adopting in every simulation is fixed, and this ratio is chosen to make that the total expense per hour on the CPU cluster and GPU cluster are nearly equivalent.

In this section, the simulation speed of the PM2D code is tested and compared with that of the frequently used PIC code of the EPOCH. Generally, a GPU has a much larger computer power than a CPU core, however, it is also more expensive than a CPU core. For the sake of fairness, the comparisons of the simulation speeds of the PM2D code (GPU-based) and PIC code (CPU-based) are carried out under a nearly equivalent cost per hour. Namely, the ratio of GPU and CPU cores in all simulations is fixed, and this ratio is inversely proportional to the cost ratio of a GPU to a CPU core per hour. On our test platform, the expense of simulation run on a GPU per hour is equivalent to that of about 80 CPU cores.

In the test simulations, different resolutions ranging from 2000 × 2000 to 10000 × 10000 are adopted. The size of the simulation domains is also enlarged with the resolution to keep the same space and time steps. All of the plasma conditions and laser properties are set to be the same for the PM2D code and the EPOCH code except that the macro particles per grid in the PM2D code is 1, while it is 100 in the EPOCH code. Each simulation is run to the time of $70T_0$, where $T_0$ is the laser period. Table 1 illustrates the total CPU time and GPU time of the EPOCHEEPOCH code and PM2D code respectively. From Table 1, one can find that under different resolutions, our PM2D code is about thirty times faster than the EPOCH code at the equivalent cost, which indicates our PM2D code is capable to simulate LPIs in large scales.

# 6 Simulation results

## 6.1 Benchmark with theory

Above all, the PM2D code is benchmarked by comparing the growth rates of SRS and SBS obtained from the PM2D simulations with the theoretical predictions. In the PM2D simulations, a rectangular region with boundary lengths of $100\lambda$ and $400\lambda$ in $x$ and $y$ directions respectively is chosen as the simulation box, where $\lambda$ is the wave length of the incident laser beam. A hydrogen plasma slab is set in the center of the simulation box with a length of $80\lambda$ and $200\lambda$ in $x$ and $y$ directions, respectively. In the simulation, the grid size is set as $dx = dy = 0.05\lambda$ and there is 1 macro particle per grid. The time step for the wave equation is $dt = 0.035$, which is selected to meet the stability condition for the numerical schemes of wave equation ($dt < \sqrt{2}/2dx$).

In the benchmark simulations, the laser beam that are injected on the left boundary of the simulation box has a Gaussian vector potential profile of $a(y) = a_0 \exp[-((y-y_0)/w)^2]$, where $y_0 = 0$ is the beam center and $w = 60\lambda$ is the characteristic width which is related to the full width at half maximum (FWHM) of a Gaussian beam as $\text{FWHM} = 2w\sqrt{\ln 2}$. The polarization of incident laser beam is set in the z-direction. Two different laser intensities of $a_0 = 0.04$ and $a_0 = 0.06$ are chosen to verify the PM2D code under different plasma densities. In the SRS simulations, the SRS growth rates are obtained for four different electron densities of $0.05n_c, 0.1n_c, 0.15n_c$ and $0.2n_c$, where $n_c$ is the critical density. The electron temperature is set as $1 keV$ and ions are assumed to be immobile in the SRS simulations. While in the SBS simulations, the electron densities are set as $0.3n_c, 0.4n_c, 0.5n_c$ and $0.6n_c$ to exclude the influences of SRS. In the SBS simulations, the electron and ion temperatures are set as $1\ keV$ and 200 eV, respectively.

In the linear theory, the growth rate of SRS can be expressed as :

$$(20) \quad \gamma_R = \frac{ka_0}{4}\left[\frac{\omega_{pe}^2}{\omega_{ek}(\omega_0-\omega_{ck})}\right],$$

where $\gamma_R$ is the growth rate of SRS that is normalized to the incident laser frequency $\omega_0$, $a_0$ is the normalized vector potential, $k$ is the wave number of electron plasma wave, $\omega_{pe}$ is the plasma frequency, and $\omega_{ek} = \sqrt{\omega_{pe}^2 + 3k^2 v_{eth}^2}$. In the linear growth stage of SRS, the amplitude of electron plasma wave grows exponentially in a growth rate given by Eq. (20), so does the scattering laser light. Based on this, the numerical growth rates of SRS in the PM2D simulations can be obtained by evaluating the growth rates of the scattering light. To this end, the time evolution of back-scattering laser light on the left boundary is firstly recorded as shown in Fig. 4(a), from which the distribution of the scattering laser light at the center line of $y = 0$ can be obtained as displayed in Fig. 4(b). If the instability of SRS grows exponentially in the linear stage, then the envelope of the scattering laser light in Fig. 4(b) should be an exponential curve before its first saturation. The logarithm of the envelop of the scattering laser light is displayed in Fig. 4(c), from which one can find at the beginning it is approximate a straight line whose slope gives the numerical growth rate of SRS in the linear stage.

The theory growth rate of SBS is given by :

$$(21) \quad \gamma_B = \frac{1}{2\sqrt{2}} \frac{k_0 a_0 \omega_{pi}}{\sqrt{\omega_0 k_0 C_s}},$$

where in Eq. (21), $\gamma_B$ is the growth rate of SBS, $\omega_0, k_0, a_0$ are the frequency, wave number and vector potential of the incident laser beam, respectively. And $\omega_{pi} = \omega_{pe}\sqrt{Zm_e/m_i}$ is the ion plasma frequency, where $C_s = \sqrt{(ZT_e/m_i)}$ is the ion acoustic velocity and $Z = 1$ is the ion charge state. The calculation of the numerical growth rates of SBS in the PM2D simulations follows a similar procedure as that for the SRS.

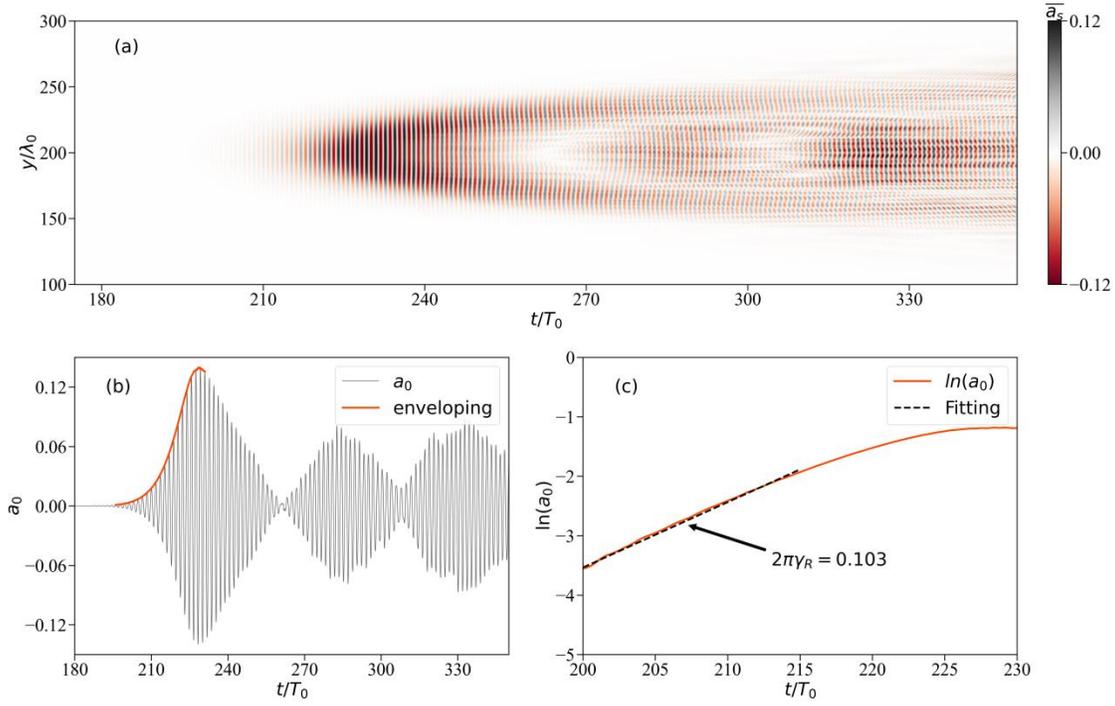

Figure 4: An example to show the calculation process of the growth rates of LPI in the PM2D code. (a) shows the time evolution of the vector potential of the scattering laser light on the left boundary of the simulation box for $0 < t < 350T_0$ in a SRS simulation, where the laser intensity is set as $a_0 = 0.06$ and the electron density is chosen as $n_e = 0.15n_c$. (b) shows the time evolution of the vector potential on the line of $y = 0$ in (a). (c) displays the logarithm of the envelop of the scattering laser light in (b) that is shown by the red line. The slope of the straight line gives the numerical growth rate.

Fig.ures 5(a) and 5(b) compare the growth rates of SRS and SBS obtained from the PM2D simulations with their theoretical predictions, respectively. It can be found that the numerical results and theoretical predictions agree well, which indicates that the PM2D code has a high precise in the simulation of both SRS and SBS.

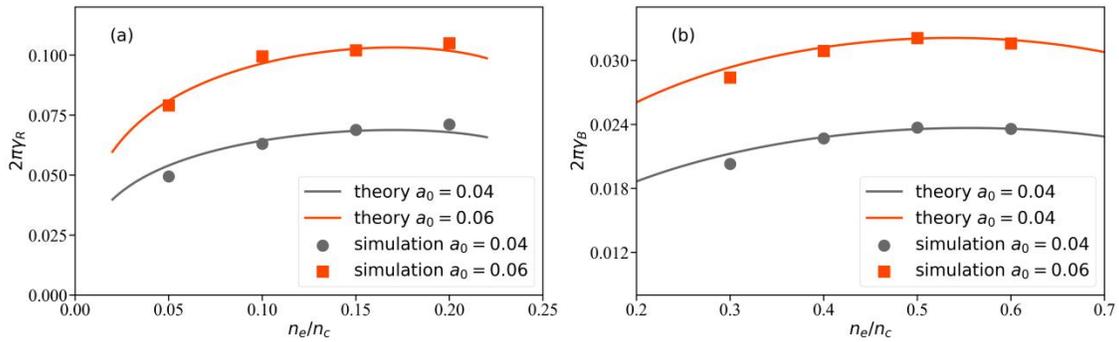

Figure 5: The growth rates of SRS (a) and SBS (b) obtained from the theoretical predictions (solid line) and PM2D simulations (discrete points) under different laser intensity (black line for $a_0 = 0.04$ and red line for $a_0 = 0.06$) as functions of the plasma density.

## 6.2 Simulation of sideward SRS

The SRS that exist in both direct drive and indirect drive ICF may not only scatter a considerable proportion of the laser energy away from the target, but also produce harmful hot electrons. In the indirect drive ICF experiments on NIF, it was found that SRS is the dominant instability for the inner cone laser beams and may result in about 20 percent of laser energy loss [8]. In the ignition-scale direct-drive experiments on NIF, it was found that SRS is the dominant instability and is responsible for the hot electron generation [14], which is different from the results in the small scale direct-drive experiments that were carried out on OMEGA, where TPD was found to be the dominant instability [46]. In this section, the typical processes of SRS are simulated by the PM2D code, and the results are compared with the PIC simulation results.

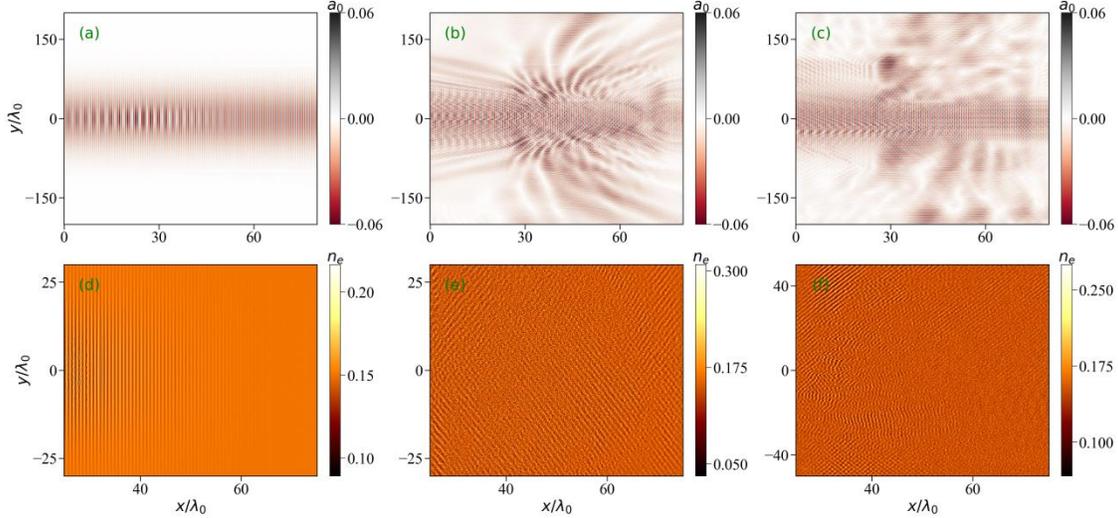

Figure 6: The snapshots of electron density (top row) and laser field (bottom row) obtained from numerical simulations, where (a) and (d) are from PM2D simulation at $t = 160T_0$, (b) and (e) are from PM2D simulation at $t = 460T_0$, and (c) and (f) are from PIC simulations at $t = 430T_0$.

In the simulation of sideward SRS, the simulation box is also a rectangle that is located within $0 < x < 100\lambda$ and $-200\lambda < y < 200\lambda$. A hydrogen plasma slab with a length of $80\lambda$ and $200\lambda$ in $x$ and $y$ directions respectively is set in the center of the simulation box. The plasma density is chosen as $n_e = n_i = 0.15n_c$ and the electron temperature is set as 1keV, the ions are assumed to be immobile to exclude the influences of SBS. A Gaussian laser beam with an expression of $a(y) = a_0 \exp[-((y-y_0)/w)^2]$ is injected from the left boundary, where $a_0 = 0.04$ is the normalized vector potential, $y_0 = 0$ is the beam center and $w = 60\lambda$ is the beam width in the $y$ direction. The space step and time step for the wave equation are chosen as $dx = dy = 0.05\lambda$ and $dt = 0.035T_0$, here $T_0$ is the laser period. The shape of a macro particle is set as a square with a length of $DD = 1.6dx$ and there is 1 macro particle per grid cell. In comparison, the EPOCH code [38] is adopted for the PIC simulations, where the simulation box, boundary conditions, initial laser-plasma conditions all keep the same as those in the PM2D simulations except that the number of macro particles per cell is set as 100.

Figures 6(a) and 6(d) show the snapshots of electron density and laser field obtained from the PM2D simulation at an early time of $t = 160T_0$. Here the laser vector potential $a_0$ is a superposed field of both the incident laser lights and scattered laser lights. From the fluctuations of electron density and laser field, one can conclude that the dominant instability is the backward SRS at the early stage [47]. In a later time of $t = 460T_0$, however, one can found from Fig. 6(b) and Fig. 6(e) that the fluctuations of electron density and laser field are no longer parallel with the $y$ axis. This implies that the sideward SRS develops and becomes the dominant instability at the later stage [48]. The development of sideward SRS at the later stage is also confirmed by the PIC simulation, as shown in Fig. 6(c) and Fig. 6(f). Moreover, obvious filamentation of the laser lights [34] are observed in both Figs. 6(b) and 6(c) that are obtained from the PM2D and PIC simulations, respectively. The laser filamentation may increase the local intensity of laser beams and result in stronger LPI excitation even when the average laser intensity is below the threshold of LPI.

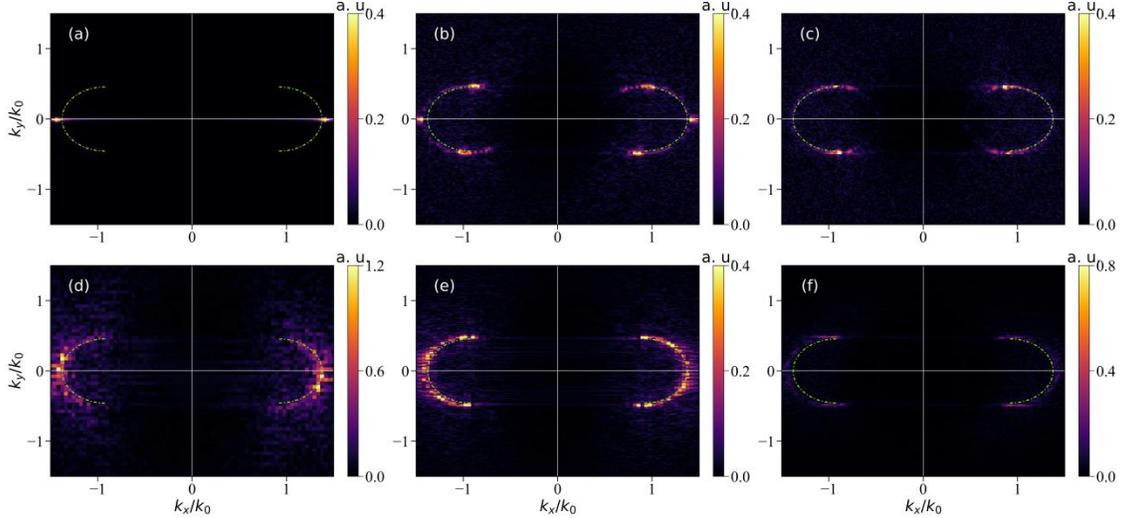

Figure 7: The distribution of wave number for the electron density fluctuations obtained by PM2D code (top row) and PIC code (bottom row) at different times, where (a), (b), (c) are obtained at $t = 160T_0, 320T_0$, and $460T_0$, respectively in the PM2D simulations, and (c), (d), (e) are obtained at $t = 130T_0, 290T_0$, and $430T_0$, respectively in the PIC simulations. the green dashed line shows the possible distribution of the theoretical wave numbers. Both of the wave numbers obtained by PM2D code an

d PIC code indicate the conversion of the back scattering SRS to the side scattering SRS.

Figure 7 shows the wave vector distributions of the fluctuating electron densities in the SRS evolution processes obtained from the PM2D and PIC simulations, respectively. In the early stage shown by Figs. 7(a), and 7(d) ($t = 160T_0$ for the PM2D code and $t = 130T_0$ for the PIC code), one can find that the wave vectors obtained by two codes both distribute around the $k_x$-axis, namely, the dominant instability is the backward SRS at this time. Figures 7(b), and 7(e) show the wave vector distributions simulated by the PM2D and PIC codes in a relatively later stage ($t = 320T_0$ for the PM2D code and $t = 290T_0$ for the PIC code), from which both of the backward and sideward scattering signals can be distinguished. This indicates the conversion of the SRS from the backward scattering to the sideward scattering. Finally, the backward SRS is almost disappeared and there is only sideward SRS at a much later time as shown in Figs. 7(c) and 7(f) simulated by PM2D code and PIC code, respectively. For the conversion from the backward to sideward SRS, the PM2D simulation results are consistent qualitatively with the PIC simulation results.

To further illustrate the conversion from the backward to sideward SRS, in Fig. 8 we draw out the time evolutions of the laser lights that propagate through the four boundaries of the PM2D simulation box. Figure 8(a) shows the backward scattering laser light within $220T_0 < t < 350T_0$, from which one can find that the scattering lights distribute in a wide range of angle after $t > 300T_0$. This indicates the onset of sideward SRS. Moreover, the time evolutions of the scattering laser lights recorded on the upper and lower boundaries indicate the development of near 90° side scattering of SRS after $480T_0$.

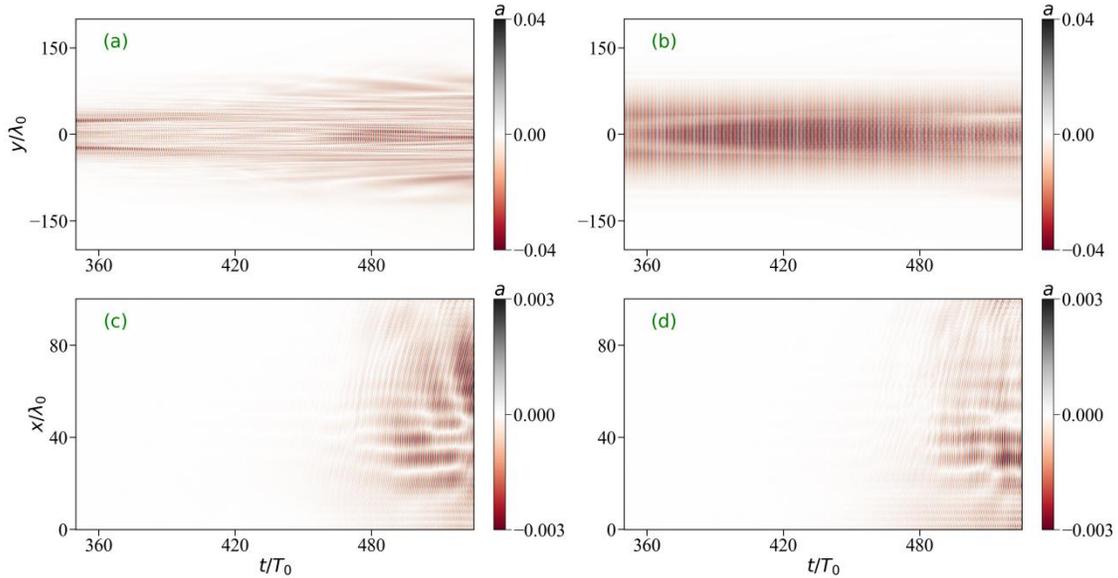

Figure 8: Time evolution of the scattering laser light on the four boundaries of the PM2D simulation box, where (a), (b), (c), (d) record the scattering laser light on the left, right, upper, and lower boundaries, respectively.

In the SRS process, electron plasma waves (EPW) will finally transfer their energies to electrons via the Landau damping effect. During this process, some electrons will be accelerated to high velocities to become the so called "hot electrons". The hot electrons are generally defined as those electrons whose energies are larger than 50 keV (or velocities larger than $0.31c$). When the velocity distribution function of electrons has a negative slop around the phase speed of EPW, the EPW will transfer its energy to the electrons and accelerate them to high velocities. This is the Landau damping process and a typical character induced by the Landau damping is that the velocity distribution function will become flat around the phase speed of EPW [35]. In Fig. 9(a), we show the electron velocity distribution simulated by the PM2D code, where the distribution function becomes flat around the phase speed of EPW [35]. For comparison, Fig. 9(b) shows the electron velocity distribution simulated by the PIC code, where a similar distribution is present. This indicates that the PM2D code is able to capture the Landau damping of EPW as well as the hot electron generation self-consistently.

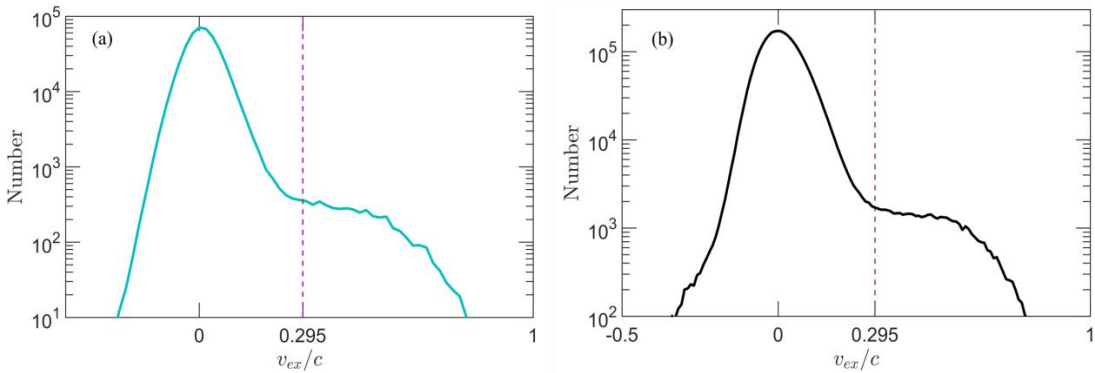

Figure 9: The velocity distributions of electrons that are located within 25λ < x < 60λ and −10λ < y < 10λ. (a) and (b) show the PM2D simulation result at t = 240$T_0$ and the PIC result at t = 310$T_0$, respectively. The purple dashed lines in (a) and (b) show the theoretical phase velocity of the electron plasma wave. The distribution functions become flat around the phase velocity, which is induced by the Landau damping effect. Both of the PM2D code and PIC code give a similar result.

## 6.3 Simulation of CBET

Cross beam energy transfer (CBET) is an important issue in the multiple-beam laser–plasma interactions in ICF. In the indirect drive ICF, it has been proved that CBET can be adopted to tune the implosion symmetry in ignition experiments that are carried out on NIF [9] [49]. In the direct drive ICF, the unabsorbed laser lights that propagate away from the target may couple with other incident laser beams to excite the CBET instability, which may induce the energy transfer from incident laser beams to the unabsorbed lights. The unabsorbed laser light will carry the transferred laser energy away from the fusion capsule, which degrades the energy deposition efficiency as well as the implosion symmetry [50]. In this section, the physical process of CBET is simulated by the PM2D code and the comparison is made with the PIC simulation.

In the PM2D simulation, the simulation box is located within $0\lambda < x < 400\lambda$ and $-200\lambda < y < 200\lambda$, where a plasma slab with a length of $360\lambda$ in both of $x$ and $y$ directions is set in the center. The plasma density is chosen as $n_e = 0.4n_c$ to exclude the influences of SRS. The electron and ion temperatures are set as 1 keV and 200 eV, respectively. Two laser beams with expressions of $a_1(y) = a_1\exp[-((y - y_1)/w_1)^2]$ and $a_2(y) = a_2\exp[-((y - y_2)/w_2)^2]$ respectively are loaded from the left boundary of the simulation box, where the normalized vector potentials of $a_1$ and $a_2$ are all set as 0.008 (corresponding to $7.12 \times 10^{14} W/cm^2$ for a 351nm laser light), the beam centers of $y_1$ and $y_2$ are set as $-50\lambda$ and $50\lambda$, respectively, and the characteristic beam widths of $w_1$ and $w_2$ are both set as $30\lambda$. Beam 1 and beam 2 are injected in angles of 15 and -15 degrees with respect to the $x$ direction and the wavelengths of beam 1 and beam 2 are set as $\lambda_0$ and $0.9995\lambda_0$, respectively, where $\lambda_0$ is chosen as 351nm. The simulation domain is divided into $8000 \times 8000$ grid cells. The time step for the wave equation is chose as $dt = 0.035T_0$, while the time step for updating the particle motion is five times larger than the time step for solving the wave equation, considering that the ion acoustic wave has a low frequency and a small growth rate. The edge length of a macro particle is set as $DD = 1.85dx$ and there is 1 macro particle per grid cell. In the PIC simulation, all of laser-plasma conditions keep the same as those in the PM2D simulation except that the number of macro particle per grid is set as 121.

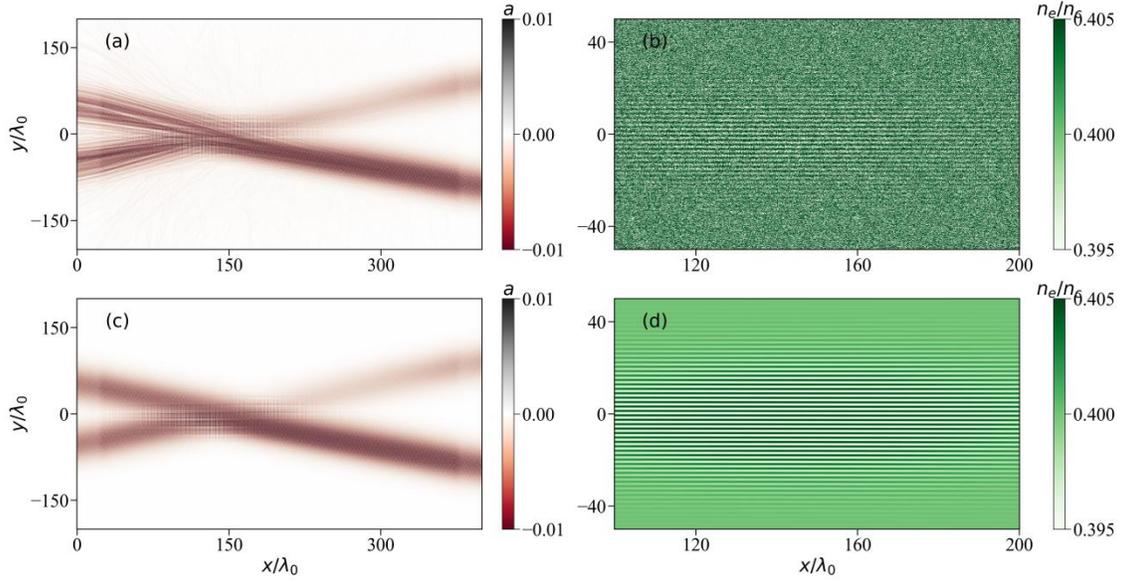

Figure 10: The snapshots of the laser fields and ion densities in the CBET process. (a) and (b) are the laser field and ion density simulated by the PIC code at $t = 875T_0$, (c) and (d) are the laser field and ion density simulated by the PM2D code at the same time.

In Fig. 10, we show the snapshots of the laser fields and ion densities simulated by PM2D code and PIC code at $t = 875T_0$. The energy transfer from the high-frequency laser beam to the low-frequency laser beam has been observed in both of the PM2D and PIC simulations as shown in Fig. 10(a) and Fig. 10(c). Due to the higher numerical noise of the PIC code, we also find that the filamentation of the incident laser beams is more serious in the PIC simulation. Figures 10(b) and 10(d) illustrate that the distributed region and spatial period of the density fluctuations (ion acoustic waves) are also consistent for the PIC and PM2D simulations.

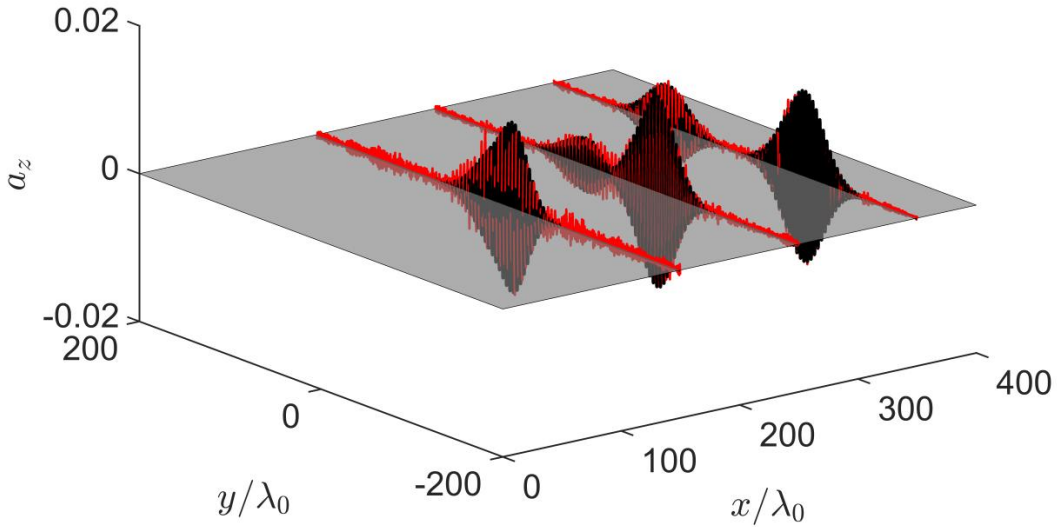

Figure 11: The transverse distributions of vector potentials ($a_z$) selected from Fig. 10(a) and Fig. 10(c) at different longitudinal posotions ( $x = 150\lambda_0, 250\lambda_0$ and $350\lambda_0$), where the red lines and black lines denote the results by the PIC code and PM2D code, respectively. The grey shading plane indicates the simulation plane.

Figure 11 compares the vector potentials that are selected from Fig. 10(a) and Fig. 10(c) at $x = 150 \lambda_0$, $x = 250\lambda_0$ and $x = 350\lambda_0$ for the PM2D and PIC simulations, respectively. It is found that the spatial laser intensity distributions simulated by two codes are in a qualitative agreement in the CBET process. It is also found that the vector potential obtained from the PIC simulation has more noise than that from the PM2D simulation, which also indicates that our PM2D code has a lower numerical noise and higher accuracy.

### 6.4 Large-scale simulation

The large spatial-temporal scale numerical simulations play a crucial role in revealing the evolution rules of LPI in ICF experiments, which helps to design effective strategies to mitigate LPI [51]. At the large scales, the laser depletion, refraction, diffraction, and filamentation may have obvious influences on the evolution of LPIs and make it distinctly different from that at the small scales. For instance, the dominant LPI mode at the large scales could be different from that at the small scales [52]. Since the evolution of LPIs is sensitive to the spatial and temporal scales, the small-scale kinetic simulations would be not competent to explain the phenomena related to the LPIs developed in ICF experiments at large scales. More importantly, it is a big challenge for the conventional PIC codes to simulate the LPIs up to several hundreds of picoseconds without numerical collapse. For example, in our 1D PIC simulations using EPOCH code, we found that the LPI signals have almost disappeared after tens of pico-seconds simulation since the electrons are heated by numerical noise to an unreasonably high temperature [39]. In this subsection, we will show that our PM code is suitable for simulating LPIs up to several hundreds of picoseconds, as well as simulating LPIs at a large scale that is relevant to the ICF experiments.

In a long-time simulation of SRS up to one hundred picoseconds, the simulation box is located within $[0,200\lambda]$ in the $x$ direction, and $[-50\lambda, 50\lambda]$ in the $y$ direction. In the center of the simulation box, a rectangular hydrogen plasma slab is set with side lengths of $180\lambda$ and $80\lambda$ in $x$ and $y$ directions, respectively. The plasma density is chosen as $n_e = 0.15n_c$, where $n_c$ is the critical density. The electron temperature is set as 1 keV and the ions are assumed to be fixed to exclude the SBS. The driver laser is injected from the left boundary, whose normalized vector potential can be expressed as $a(y) = a_0 \exp[-(y/w)^2]$ with $a_0 = 0.02$ (corresponding to $4.45 \times 10^{15} W/cm^2$ for a 351 nm laser light) and the characteristic beam width $w = 30\lambda$. The resolution is chosen as $4000 \times 2000$, and the time step for solving the wave equations is $dt = 0.035T_0$. The shape of the macro particle is chosen as a square with a side length of $DD = 1.9dx$ and 1 macro particle is set in every grid. The PM2D code was run up to three million time steps (about 120 picoseconds for a central laser wavelength of 351 nm) to test its numerical stability.

In Fig. 12(a), we show the time evolution of scattering laser energy that flow out the left boundary of the simulation box. Here, the scattering laser energy is time-averaged over every 0.2ps. Figure 12(a) illustrates that the back scattering laser lights are always active up to the end of the simulation at t=120 ps, which demonstrates the robustness of our PM2D code in the long-time simulation. In Fig. 12(b), we show the time evolution of the scattering laser light recorded on the left boundary at the final simulating stage of [122.65ps, 122.85ps]. It is found that the wave front of the scattering laser light becomes crooked and disrupt, which may be induced by the Gaussian type distribution of incident laser intensity. According to the electron density and temperature used in the simulation, the frequency of the back-scattering laser light can be theoretically predicted as $\omega_s = 0.67\omega_0$, where $\omega_0$ is the frequency of the incident laser light. As shown in Fig. 12(c), the frequent spectrum of the back-scattering laser light obtained from the PM2D simulation peaks around $\omega = 0.67\omega_0$, which is in good quantitative agreement with the theoretical prediction. Figure 12(d) shows the transmitted laser light recorded on the right boundary, and the corresponding frequency spectrum displayed in Fig. 12(e) indicates that the forward scattering is very weak and the frequency peaks at $\omega_0$. The above simulation results illuminate that our PM2D code is competent for the long-time simulation of LPIs.

After PM2D code is transplanted onto the GPU cluster, it can be used for large spatial scale simulations in which the total number of mesh grids can be as large as several billions. To illustrate this, we have carried out a large-scale simulation on the GPU cluster with 16 GPU nodes, where each GPU node has four A100 GPUs, i.e., the total number of GPUs is 64. In this simulation, the simulation box has a length of 2500 $\lambda$ in $x$ direction and 1200 $\lambda$ in $y$ direction. The resolution of the simulation is set as $50001 \times 24001$, with a total mesh grids of over 1 billion. To simulate the coronal plasma conditions as obtained in the Omega laser facility, the plasma density varies exponentially in the $x$ direction as $n_e(x) = n_0 \exp[(x - x_1)/L_n]$, where $n_0$ is chosen as $0.05nc$, $x_1$ is the start point of the plasma lab, $L_n = 237.4$um is the scale length of the plasma density. To mitigate the thermal expansion, we set the density slopes at the two edges of the plasma slab in the $y$ direction. The temperature of electrons and ions are 1 keV and 500 eV, respectively. The incident laser beam is a Gaussian beam with an intensity profile of $a(y) = a_0 \exp[-((y - y_0)/w)^2]$, where $a_0$ is set as 0.01 (corresponding to an intensity of $1.11 \times 10^{15}$ W/cm$^2$ for a laser wavelength of 351 nm ). The laser beam is obliquely injected on the left boundary of the simulation box with an angle of $\theta = 10°$. The central point of the incident laser is chosen as $y_0 = 200\lambda$, and the characteristic beam widths is chosen as $w = 70\lambda$. The time step for solving the wave equations is chosen as $dt = 0.035T_0$. The shape of the macro-particle is chosen as a square with a side length of $DD = 1.9dx$ and 1 macro particle is set in every grid cell. We run the PM2D code up to 5.2 ps ($4500T_0$), which costs a total computation time of around 12 hours.

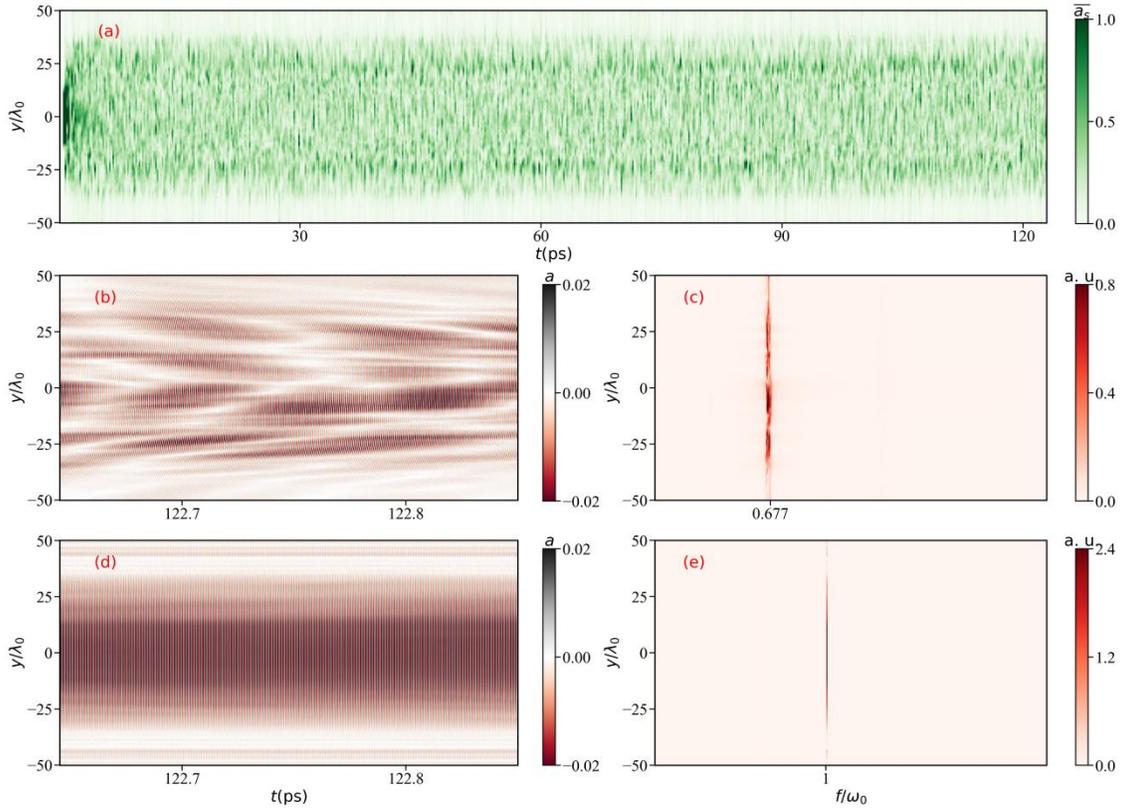

Figure 12: Numerical stability test of the PM2D code run up to 120ps for an incident laser wavelength of 351 nm. Here (a) shows the time evolution of scattered laser energy that flow through the left boundary of the simulation box. (b) and (d) show the time evolution of the scattering laser lights on the left and right boundaries of the simulation box within [122.62ps, 122.85ps], respectively. (c) and (e) are the spatial distributions of frequency spectra in the $y$ direction of the laser lights in (b) and (d), respectively.

In Fig. 13, we show the snapshots of the laser fields in the large-scale simulation at $t = 3.68$ ps and 4.5 ps, respectively. It can be found that at $t = 3.68$ ps the LPIs just start to develop and thus the propagation path of the incident laser beam can be clearly recognized. As theoretically predicted, Fig. 13(a) also illustrates that the propagation of the incident laser beam is stopped when $n_e = \cos^2\theta\, n_c \simeq 0.97 n_c$. Moreover, the bending of the incident laser beam can be distinguished in Fig. 13(a), which is induced by the refraction of laser beam in an inhomogeneous plasma. At $t = 4.5$ ps, the LPIs are sufficiently developed as shown in Fig. 13(b). As a result, the 'explosion' of the incident laser beam appears, accompanied by the sideward LPIs. The sideward LPIs tends to disperse the laser energy into a wider region, which may enhance the efficiency of the laser energy absorption but also degrade the implosion symmetry.. In Figs. 13(c) and 13(d), we zoom in the vector potential distribution in the dashed rectangular area of Figs. 13(b) and 13(c), respectively, from which the excitation of SBS as well as the filamentation of the incident laser beam can be distinguished clearly.

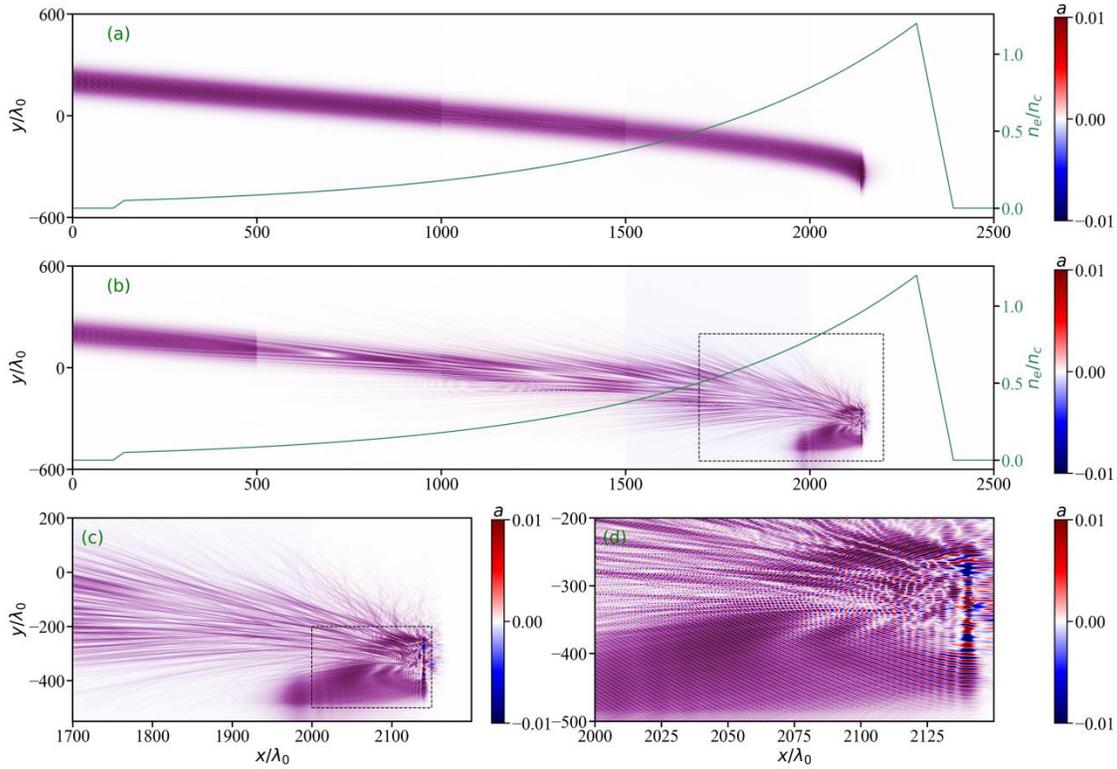

Figure 13: The snapshots of the laser field in the large-scale simulation of LPIs using PM2D code at (a) t=3.68ps when LPIs are not sufficiently developed, and (b) at t=4.5 ps when LPIs are sufficiently developed. Because the grid number in this simulation is very huge, the drawing of the snapshot of the laser field as a whole figure is hard for the typical mapping software. Here, the snapshots of laser fields are stitched by five smaller figures. The blue line in (a) and (b) shows the electron density distribution in the $x$ direction. (c) and (d) show the closeups of the laser fields in the dashed rectangular regions of (b) and (c), respectively.

To further find what is the dominant instability mode at $t = 4.5$ps when the LPIs are sufficiently developed, we display the electron density fluctuations and the corresponding wavenumber spectra in two different subdomains in Fig. 14. The first subdomain is chosen as $[750\lambda, 800\lambda] \times [0,100\lambda]$ in the $x - y$ plane, which corresponds to the relatively low density region ranging from $n_e = 0.12n_c$ to $0.129\ n_c$. From Fig. 14(a), one can find that many parallel density peaks of the plasma wave. The incident laser beam will couple with the plasma wave to produce the scattering laser light. Figure 14(b) shows the wavenumber spectrum of the plasma wave displayed in Fig. 14(a), which indicates that the SRS is the dominant instability mode in this relatively low density region at $t = 4.5$ps. Figure 14(c) shows the electron density fluctuation in the second subdomain that ranges from $x = 1780\lambda$ to $1920\lambda$, and from $y = -220\lambda$ to $-120\lambda$. The second subdomain corresponds to the relatively high density region ranging from $n_e = 0.5284n_c$ to $0.646n_c$. From Fig. 14(c), one can also find the obvious plasma wave in this high density region. Since the electron density is larger than the quarter critical density, the dominant instability is the SBS, which is confirmed by the wave number spectrum shown in Fig. 14(d). More interestingly, the filamentation of the plasma wave appears in Fig. 14(c), which correlates with the filamentation of the incident laser light as shown in Fig. 13(b).

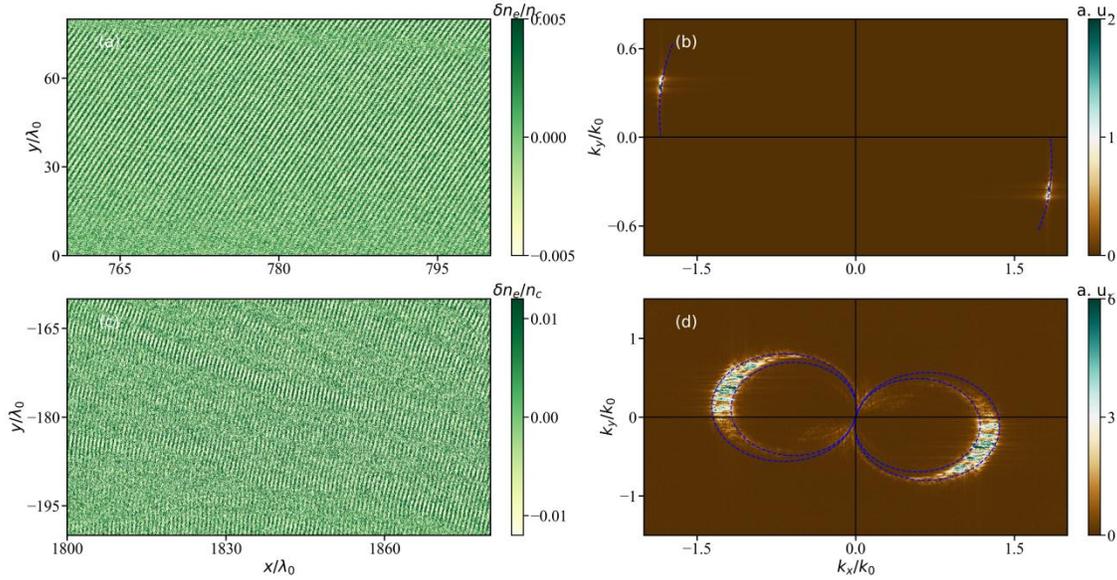

Figure 14: Closeups of the fluctuating electron density at $t$ = 4.5ps when LPIs are sufficiently developed in (a) the relatively low density region with $n_e$ ranging from $0.12n_c$ to $0.129\ n_c$, and (c) the relatively high density region with $n_e$ ranging from $n_e = 0.5284n_c$ to $0.646n_c$. (b) and (d) are the wavenumber spectra of the electron density fluctuations in (a) and (c), respectively. The blue dashed lines in (d) show the possible wave number distribution for different scattering directions of SBS under two densities of $0.5284n_c$ and $0.646n_c$, from which one can distinguish the development of side scattering SBS.

## Summary and discussion

In this paper, a two-dimensional full wave fluid model that describes the evolution of laser plasma instabilities including SRS, SBS, CBET, and laser filamentation is constructed, which is numerically solved by the particle-mesh (PM) method to gestate a cutting-edge PM2D code. As the physical model of the PM2D code is developed specifically for the LPI simulations, the required macro-particles per grid can be greatly reduced (so does the computational cost) comparing with conventional PIC codes that usually require a large number of macro-particles per grid to simulate the LPIs. Moreover, since the macro-particles can move freely across the meshes, the PM2D code can capture the kinetic effects such as the Landau damping and hot electron generation self-consistently as the conventional PIC code.

To further speed up the PM2D code, we transplant it onto the GPU platform. Since a single GPU card has the equivalent compute power of $O(1000)$ CPU cores, the GPU-based PM2D code runs dramatically faster than that based on CPU. Using the MPI, the PM2D code is parallelized on the GPU cluster using hundreds of GPUs, which makes it competent for the kinetic simulations of LPIs at the ICF experimental scales with billions of meshes. The large-scale kinetic simulations play a unique role in studying the complex LPI processes in ICF, which is a precondition for the control or mitigation of the LPIs.

Our PM2D code is benchmarked by the theory. The growth rates of SRS and SBS obtained from the PM2D simulations are in a quantitative agreement with the theoretical predictions for different laser intensities and plasma densities. To the best of our knowledge, it is the first kinetic simulation code that is quantitatively benchmarked for the simulations of LPIs. Further, the capacities of our PM2D code in simulating multi-dimensional LPIs and multi-beam LPIs are respectively exemplified by the simulations of the sideward SRS and the CBET, in which the kinetic effects such as the Landau damping and hot electron generation are also treated properly.

The robustness of our PM2D code has been verified by the simulations at large spatial-temporal scales. A reasonable scattering of laser light with the predicted frequency is observed in the PM2D simulation of the SRS up to 122 ps. In contrast, the plasma may be heated by strong numerical noise to an unreasonably high temperature after tens of pico-seconds in the typical PIC simulations of LPIs. Within a simulation domain of $1200\lambda \times 2500\lambda$, the development of various kinds of LPIs at large scales is simultaneously studied by the PM2D simulation using billions of meshes and macro-particles. It is found that the development of parametric instabilities such as SBS and SRS is strongly coupled with other processes in laser-plasma interactions such as laser depletion, hosing, and filamentation. Fortunately, our PM code is competent to simulate the development of parametric instabilities in laser-plasma interactions at large spatial-temporal scales, which is of great significance to the study of LPIs in ICF experiments.

In the next studies, we may concentrate on the nonlinear effects in the evolution of LPIs at large spatial-temporal scales and try to find the effective strategies to mitigate or control the LPIs. It should be pointed out that our PM2D code has not included the TPD instability yet, the difficulties mainly come from how to couple the TPD in the physical model given by Eqs. (1)-(3).

# Acknowledgements

This work is supported by the National Natural Science Foundation of China (Grant No. 12135009), the Strategic Priority Research Program of Chinese Academy of Sciences (Grant No. XDA25050100 and XDA25010100), Natural Science Foundation for the Youth of China (Grant No. 12305259), China Postdoctoral Science Foundation (Grant No. 2022M722109).